\documentclass[%
 amsmath,amssymb, reprint,
 aps, longbibliography
]{revtex4-1}

\usepackage{graphicx}% Include figure files
\usepackage{dcolumn}% Align table columns on decimal point
\usepackage{bm}% bold math
\usepackage{tipa}

\usepackage{siunitx}
\usepackage{xcolor, soul}
\sethlcolor{white}

\usepackage{mathrsfs}

\begin{document}

\title{Spontaneous Coulomb fissions of drops on lubricated surfaces}

%%%%%%%%%%%%%%%%%%%%%%%%%%%%%%%%%%%%%%%%%%%%%%%%%%%%%%%%%%%%%%%%%%%%%
%\newcommand*\mycommand[1]{\texttt{\emph{#1}}}
%\usepackage{xcolor, soul}
%\sethlcolor{white}
%% Each author should be given as a separate \author command.
%%
%% Corresponding authors should have an e-mail given after the author
%% name as an \email command. Phone and fax numbers can be given
%% using \phone and \fax, respectively; this information is optional.
%%
%% The affiliation of authors is given after the authors; each
%% \affiliation command applies to all preceding authors not already
%% assigned an affiliation.
%%
%% The affiliation takes an option argument for the short name.  This
%% will typically be something like "University of Somewhere".
%%
%% The \altaffiliation macro should be used for new address, etc.
%% On the other hand, \alsoaffiliation is used on a per author basis
%% when authors are associated with multiple institutions.
%%%%%%%%%%%%%%%%%%%%%%%%%%%%%%%%%%%%%%%%%%%%%%%%%%%%%%%%%%%%%%%%%%%%%

\author{Marcus Lin$^{1,2}$}
    \altaffiliation{M. L. and P. Z. contributed equally to this work}
\author{Peng Zhang$^{1}$}
    \altaffiliation{M. L. and P. Z. contributed equally to this work}
\author{Aaron D. Ratschow$^{3}$}
\author{Oscar Li$^{1}$}
\author{Sankara Arunachalam$^{1}$}
\author{Dan Daniel$^{1,2}$}
    \email{dan-daniel@oist.jp}

\affiliation{$^{1}$Droplet Lab, Physical Science and Engineering (PSE) Division, King Abdullah University of Science and Technology (KAUST), Thuwal 23955-6900, Saudi Arabia}
\affiliation{$^{2}$Droplet and Soft Matter Unit, Okinawa Institute of Science and Technology Graduate University, Onna, Okinawa 904-0495, Japan}
\affiliation{$^{3}$Max Planck Institute for Polymer Research, 55128 Mainz, Germany}

\begin{abstract}
Charged water drops are more widespread than commonly acknowledged. For example, raindrops typically carry charges of order $Q \sim$ 1 pC, while routine pipetting in the laboratory produces drops with $Q \sim$ 50 pC. Here, we show that such modest charging can spontaneously generate periodic Coulomb fissions for evaporating water drops on lubricated surfaces, with more than 60 successive cycles observed over 30 min. Interestingly, the underlying instability can be quantitatively predicted by two fissility thresholds: one marking the onset of drop elongation and another triggering fission. Each fission culminates with a fine liquid jet that disintegrates into 40--50 microdroplets, expelled within microseconds. The phenomenon spans an extraordinary range of length scales (from millimetres to microns) and time scales (hour to microseconds), with broad potential applications ranging from nanoscale fabrication to electrospray ionization.
\end{abstract}
\maketitle

Various natural phenomena and industrial processes involve the evaporation of charged drops. Examples include thunderstorm electrification \cite{gaskell1977airborne, feynman1964electricity}, inkjet printing \cite{park_high-resolution_2007}, and electrosprays \cite{gomez_charge_1994, fenn1989electrospray}. In 1882, Lord Rayleigh established the stability criterion for electrically charged drops \cite{rayleigh_xx_1882}. An evaporating drop with charge $Q$ becomes unstable upon reaching the critical radius $R_\text{c}$:
\begin{equation} \label{eq:Rayleigh}
Q = 8\pi \sqrt{\gamma \varepsilon_\text{o}}\, R_\text{c}^{3/2},
\end{equation}
now known as the Rayleigh limit; $\gamma$ is the surface tension and $\varepsilon_\text{o}$ the permittivity of free space. The same Rayleigh limit can be recast in terms of the dimensionless fissility,
\begin{equation} \label{eq:X}
X = \frac{Q^{2}}{64 \pi^{2} \gamma \varepsilon_{0} R^{3}},
\end{equation}
with instability occurring at $X_{\text{R}} = 1$ \cite{duft2003rayleigh, li_charge_2005}.

For decades, Rayleigh's prediction remained largely theoretical. Experimental confirmation only began in the 1960s \cite{doyle_behavior_1964, abbas_instability_1967}, and since then the Rayleigh limit has been extensively validated \cite{taylor_disintegration_1964, basaran_axisymmetric_1989, fernandez_de_la_mora_outcome_1996, achtzehn_coulomb_2005, taflin_electrified_1989, davis_rayleigh_1994, grimm_dynamics_2005}, but almost exclusively for \textit{levitated} drops (typically tens of microns in size) \cite{gomez_charge_1994, duft_shape_2002, grimm_field-induced_2003, singh_subcritical_2021}. For such isolated drops, studies consistently reveal a single fissility threshold $X_{\text{R}} = 1$ (corresponding to $R_\text{c}$) that governs the onset of \textit{both} elongation and fission; once $X_{\text{R}} > 1$, the initially spherical drop rapidly elongates and, within $\sim$ \SI{100}{\micro\second}, undergoes Coulomb fission, producing jets to expel excess charge \cite{duft2003rayleigh}. 

Remarkably, no Coulomb fission has been reported for \textit{sessile} drops evaporating on surfaces \cite{wilson2023evaporation, chen_evaporation_2012, guan_evaporation_2015, charitatos_droplet_2021, erbil_droplet_2023}, even though such drops are likely to be charged through routine laboratory handling, e.g., pipetting which charges water drops through contact electrification \cite{choi2013spontaneous, wang2019origin, nauruzbayeva2020electrification, ratschow2025liquid}. The absence of fission can be attributed to contact-line pinning, which constrains drop elongation and also facilitates charge loss \cite{lin2020quantifying, li2022spontaneous, singh2025bipolar, ratschow2025liquid}. 

Here, we show that \textit{millimetric} water drops on a plastic petri dish coated with a nanometric silicone oil film (i.e., a lubricated surface \cite{wong2011bioinspired, lafuma2011slippery}) spontaneously undergo periodic Coulomb fissions, exceeding 60 cycles over 30 min \cite{Lin2025exploding}. The addition of silicone oil is essential, as it eliminates contact-line pinning \cite{daniel2017oleoplaning}.  Each cycle proceeds in two stages: elongation, followed by fission. Each fission partially discharges the drop (2\% charge loss), causing contraction, while continued evaporation reconcentrates charge, driving renewed elongation and triggering the next fission, thus sustaining the oscillation cycles.

The electrical instability in sessile drops is also richer than Rayleigh's classical results. Instead of a single fissility threshold, we identify two: $X_{\text{e}} = 0.25$ and $X_{\text{c}} = 0.26$, corresponding to two distinct critical radii $R_{\text{e}}$ and $R_{\text{c}}$, which govern the onset of elongation and Coulomb fission separately. By contrast, $R_{\text{e}} = R_{\text{c}}$ for levitated drops. Our work therefore highlights the qualitatively different nature of Coulomb fission in a sessile geometry.

We further find that lubricant viscosity strongly influences the size of ejected microdroplets. Understanding the onset of fission and factors controlling the progeny droplet size may open opportunities for applications ranging from fabrication of nanomaterials \cite{park_high-resolution_2007, feng2014nanoemulsions, keiser2017marangoni, montanero2020dripping} to electrospray ionization of large molecules \cite{gomez_charge_1994, fenn1989electrospray}.

\section*{Results and discussions}
\begin{figure*}[!htb]
\centering
\includegraphics[scale=1]{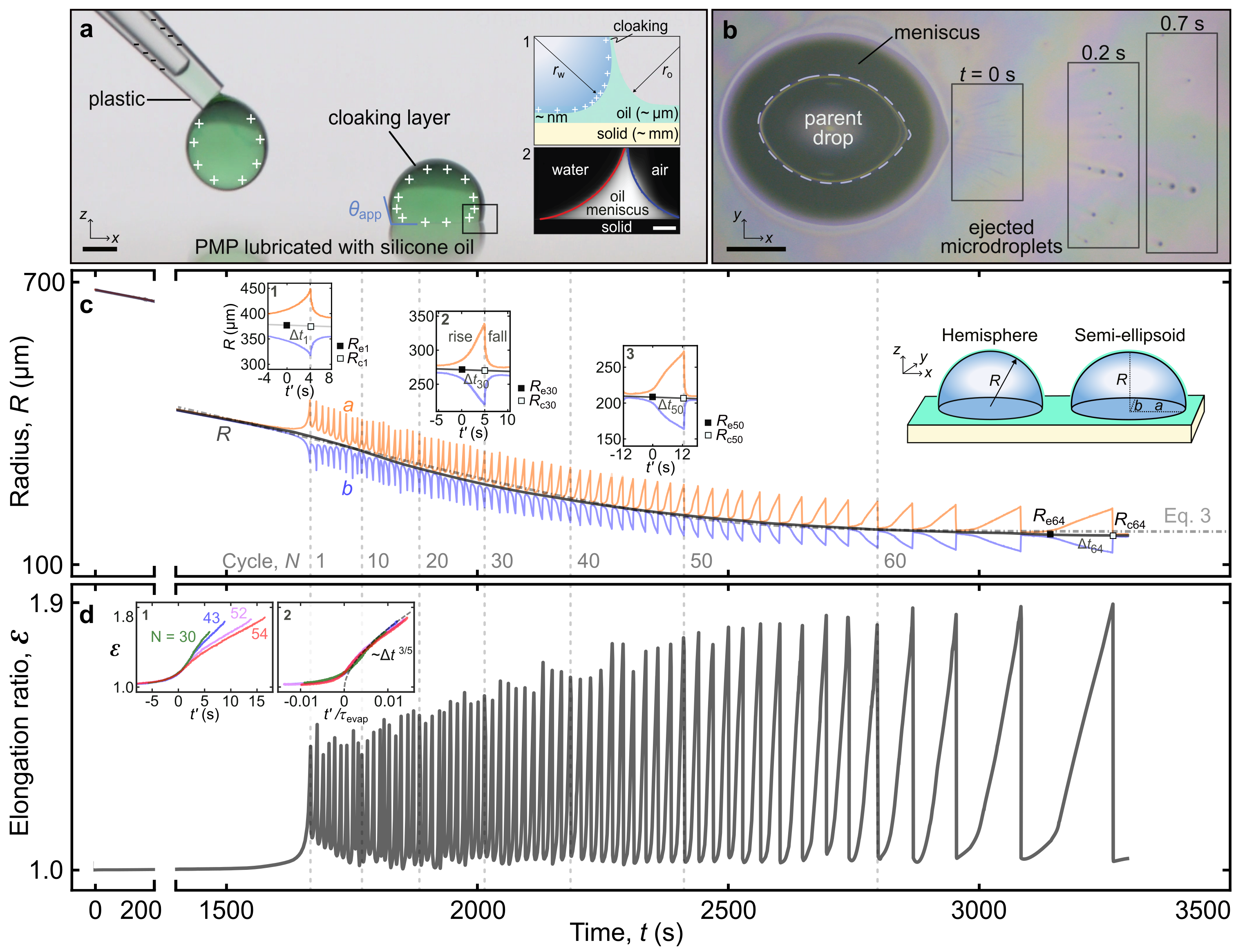}
\caption{\label{fig:fission} \textbf{Evaporation drives periodic Coulomb fissions and drop oscillations}. (a) Pipetting charges water drop; food dye added here (only) for clarity. Scale bar: 1\,mm. Insets 1, 2: schematic and confocal microscopy of the meniscus skirt (Scale bar: \SI{25}{\micro\meter}). (b) Spontaneous fission triggered by evaporation. Insets: superimposed images of ejected microdroplets. Scale bar: \SI{200}{\micro\meter}. (c) Temporal evolution of drop mean radius $R$ (solid gray line), showing periodic oscillations following fissions. $a$ and $b$ are the semi-major and semi-minor axes, while dashed-dot line is solution to Eq.~\ref{eq:dRdt}. Insets~1--3: Zoom-ins of the oscillations for $N = 1$, $30$, and $50$;  $t'$ is relative time, defined such that $t'=0$ s marks the onset of elongation. (d) Corresponding elongation ratio $\mathcal{E} = a / b$ vs. $t$. Insets 1, 2: rise phase of $\mathcal{E}$ vs. $t'$ and  $t'/\tau_{\text{evap}}$ for different $N$. }	
\end{figure*}

We used a conventional micropipette to deposit a \SI{1}{\micro\liter} deionized (DI) water drop (radius, $R$ = \SI{680}{\micro\meter}) onto a commercially available polymethylpentene (PMP, thickness: 1 mm) petri dish lubricated with a \SI{0.5}{\micro\meter} thick silicone oil film (viscosity $\eta = \SI{10}{\milli\pascal\second}$). Through contact electrification, the drop acquired an initial charge of $Q_\text{i} = +70~\text{pC}$, measured using a Faraday cup and an electrometer (Materials and Methods).  Even though other charging methods exist \cite{taylor_disintegration_1964, wang2020facile, yu2025charged}, the choice of pipetting is deliberate---it serves to highlight the surprising influence that routine laboratory handling can have on classical evaporation experiments. 

The drop behaves as a conductor, with all charges residing on its surface (Fig.~\ref{fig:fission}a). Evaporation then concentrates the charge until the electrostatic repulsion surpasses the surface tension, spontaneously triggering the first Coulomb fission approximately 20 minutes later, with jetting and microdroplet ejection (Fig.~\ref{fig:fission}b; supplementary Video \hl{S1}). All experiments were performed under controlled ambient temperature $T = 19.5 \pm 0.5^{\circ}\text{C}$ and relative humidity $RH = 63 \pm 2\%$.  

Because silicone oil fully wets water, the millimetric drop is cloaked in a micron-thick oil film, resulting in an effective surface tension of $\gamma_\text{eff} = \gamma_\text{oa} + \gamma_\text{ow} = \SI{60}{\milli\newton\per\meter}$, where $\gamma_\text{oa} = \SI{20}{\milli\newton\per\meter}$ is the oil surface tension and $\gamma_\text{ow} = \SI{40}{\milli\newton\per\meter}$ is the oil--water interfacial tension (Fig.~\ref{fig:fission}a) \cite{kreder2018film}. At the same time, an oil meniscus skirt with radius of curvature $r_{\text{o}} = \SI{80}{\micro\meter}$ wraps around the drop (blue curve, Inset 2, Fig.~\ref{fig:fission}a) \cite{kreder2018film, semprebon2021apparent, dai_droplets_2022}, forming a low-pressure region of magnitude $\gamma_{\text{oa}}/r_{\text{o}}$ at the base. To balance this Laplace pressure, the bottom of the drop bulges outward with a curvature radius $r_{\text{w}} = (\gamma_{\text{ow}}/\gamma_{\text{oa}}) r_{\text{o}} = \SI{135}{\micro\meter}$ (red curve, Inset 2). Since $r_{\text{w}} \ll R$, the meniscus amplifies the local electric field and charge density by a factor of $R/r_{\text{w}} \sim 5$ (Inset 1, Fig.~\ref{fig:fission}a).  

Although the drop appears to have a finite contact angle $\theta_{\text{app}} = 110^{\circ}$ (Fig.~\ref{fig:fission}a), this angle is only apparent. In reality, a continuous nanometric lubricant film lies beneath the drop, stabilised by repulsive van der Waals interactions (arising from the specific material combination of PMP and silicone oil), meaning that a true three-phase contact line does not exist (Inset 1; Supplementary Fig.~\hl{S1} and Video \hl{S2}) \cite{daniel2017oleoplaning}. The complete elimination of contact-line pinning is key to observing fissions.

Macroscopically, the drop geometry is well approximated by a hemisphere. For diffusion-limited evaporation, its mean radius $R$ decreases as  
\begin{equation} \label{eq:dRdt} 
\frac{\mathrm{d}R}{\mathrm{d}t} = - \frac{D \Delta c}{\rho R} \left(1 - \frac{\ell_{0}}{R} \right), 
\end{equation}
where the term $(1 - \ell_{0}/R)$ accounts for the meniscus of characteristic size $\ell_{0} \sim r_{0}$, which impedes vapour transport \cite{guan2015evaporation}. Without this term, Eq.~\ref{eq:dRdt} reduces to the classical $R^{2}$-law \cite{gelderblom2022evaporation}. Here, $D = \SI{2.4e-5}{\meter\squared\per\second}$ is the diffusion constant of water vapour, $\rho = \SI{998}{\kilogram\per\meter\cubed}$ is the water density, and $\Delta c = (1 - RH)c_{\text{s}}(T)$ is the undersaturation relative to the saturated concentration $c_{\text{s}}(\SI{19.5}{\celsius}) = \SI{1.7e-2}{\kilogram\per\meter\cubed}$ at the interface. The cloaking layer is assumed too thin to hinder diffusion \cite{bian2021rediscovering}.

Numerically integrating Eq.~\ref{eq:dRdt} using $\ell_{0} = \SI{163}{\micro\meter}$ as the only fitting parameter (dashed--dot line in Fig.~\ref{fig:fission}c) shows excellent agreement with the experiment (solid line). As the drop shrinks and $R$ approaches $\ell_{0}$, evaporation slows down---sharply contrasting with the classical $R^{2}$-law. Notably, the model continues to apply even after fissions start and drop oscillations begin ($t > 1600$~s). 

Each oscillation cycle $N$ consists of two stages: elongation and fission. Elongation begins at the critical radius $R_{\text{e}N}$, and Coulomb fission only occurs when the drop shrinks further to $R_{\text{c}N} < R_{\text{e}N}$. The small offset between $R_{\text{e}N}$ and $R_{\text{c}N}$---typically only a few microns---introduces a measurable delay $\Delta t_{N}$ between the onset of elongation and subsequent fission (Insets, Fig.~\ref{fig:fission}c).  

For example, in the first cycle, elongation occurs at $t = \SI{1662}{\second}$ when the drop reaches $R_{\text{e1}} = \SI{377}{\micro\meter}$, and fission follows $\Delta t_{1} = 4$~s later when the drop further shrinks to $R_{\text{c1}} = \SI{375}{\micro\meter}$ (Inset~1, Fig.~\ref{fig:fission}c). The existence of two critical radii and a significant time delay between elongation and fission are unique to sessile drops; by contrast, levitated drops exhibit only a single critical radius (Eq.~\ref{eq:Rayleigh}), with elongation and fission unfolding within $\sim$ \SI{100}{\micro\second} of each other \cite{duft_shape_2002}.

During the interval $\Delta t_{1}$, the drop gradually elongates into a semi-ellipsoid, reaching a peak elongation ratio $\mathcal{E}_{\text{peak}} = a/b = 1.4$, where $a$ and $b$ are the semi-major and semi-minor axes. Simultaneous top- and side-view images confirm that the oscillations are confined to the $x$--$y$ base plane (Supplementary Fig.~\hl{S2} and Video \hl{S3}), driven by the meniscus-amplified electric field (schematic in Fig.~\ref{fig:fission}c). Fission then occurs, producing a jet that partially discharges the drop and causing contraction, i.e., $\mathcal{E} \rightarrow 1$. Further evaporation reconcentrates charge, driving renewed elongation and fission, thus sustaining over 60 cycles over the next 30~min ($t = 1650$--3300 s).  

\begin{figure}[!htb]
\centering
\includegraphics[scale=1]{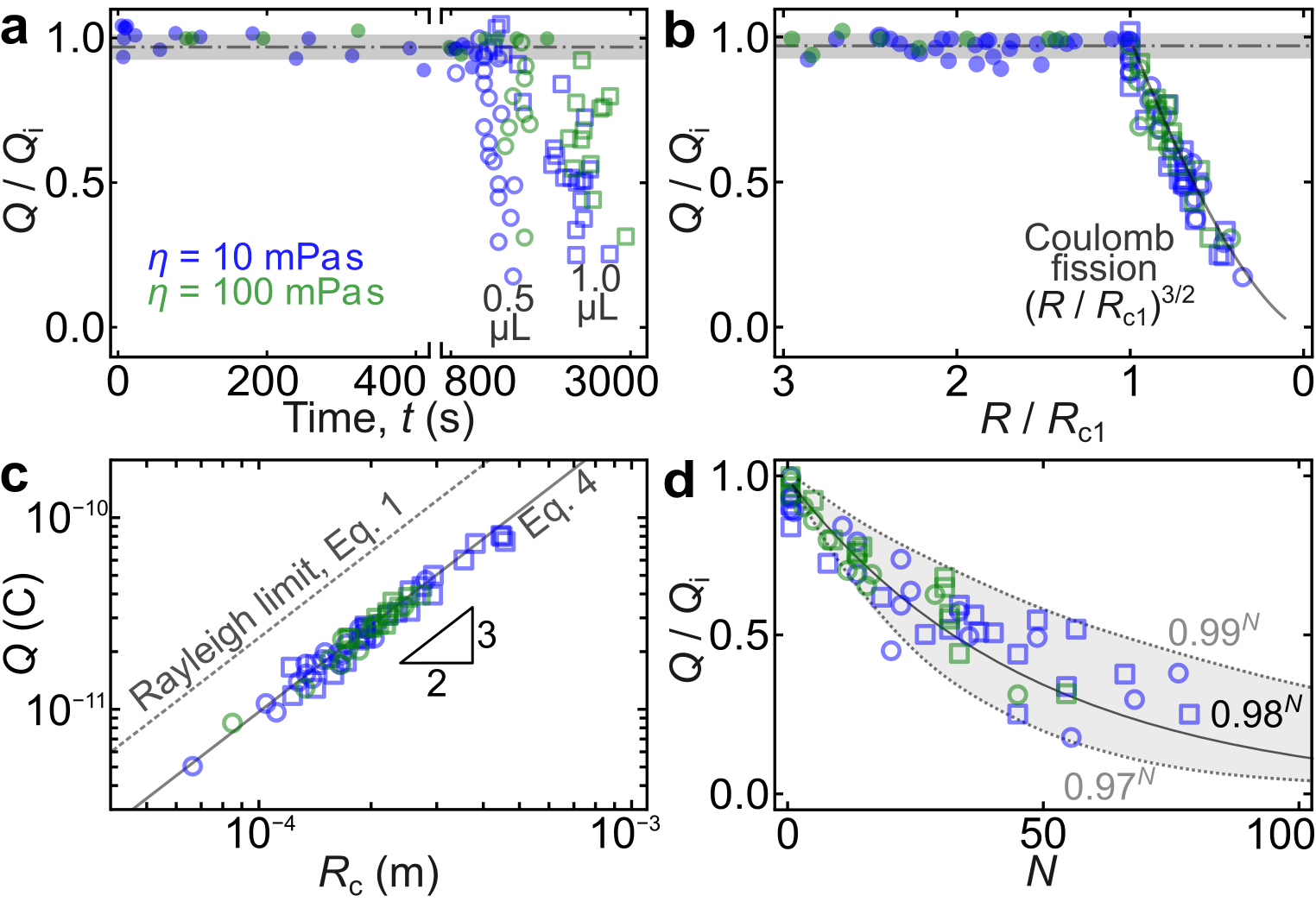}
\caption{\label{fig:discharge} \textbf{Universal charge decay during fissions.} Fractional charge $Q/Q_\text{i}$ as a function of (a) $t$ and (b) $R/R_\text{c1}$. The charge is conserved without fissions (filled markers). Once fission starts (open markers), charge decay follows the universal relation $Q/Q_\text{i} = (R/R_\text{c1})^{3/2}$. (c) The same data is consistent with Eq.~\ref{eq:Q_crit_prefactor}, and (d) charge decays as $Q/Q_\text{i} = 0.98^{N}$ (solid lines). Circular and square markers denote 0.5 and \SI{1.0}{\micro\liter} drops, while different lubricant viscosities are distinguished by colour. Measurements are from 102 individual drops.}
\end{figure}

The evaporating drop reaches progressively smaller critical radii, down to $R_{\text{e50, c50}} = 208, \SI{206}{\micro\meter}$ for $N$ = 50 and  $R_{\text{e64, c64}} = 168, \SI{165}{\micro\meter}$ for $N$ = 64. As the evaporation rate slows, $\Delta t_{N}$ grows with $N$, up to $\Delta t_{50}$ = 12 s and $\Delta t_{64} = 120$~s (Insets, Fig.~\ref{fig:fission}c). The peak elongation ratio $\mathcal{E}_{\text{peak}}$ likewise increases from 1.4 to 1.9, showing that smaller drops are more elongated (Fig.~\ref{fig:fission}d, Supplementary Fig.\hl{~S3}). 

The origins of the oscillation dynamics and of two critical radii will be explained later. For now, it suffices to note that $\Delta t_{N}$ and hence the rise times are set by the characteristic evaporation time $\tau_{\text{evap}} = R / \lvert \mathrm{d}R/\mathrm{d}t \rvert$.  When the rise phases of $\mathcal{E}$ for $N = 30$--54 are rescaled by $\tau_{\text{evap}}$, they collapse onto a single master curve (Insets~1 and~2, Fig.~\ref{fig:fission}d). In contrast, the fall times are determined, at least partly, by the inertial-capillary timescale $\tau_{\gamma}$ = $(\rho R^{3}/\gamma_\text{eff})^{1/2} \sim 1$ ms. The smooth return to a hemisphere without oscillations indicates an overdamped relaxation.
\bigskip

\textbf{Modified Rayleigh limit}. To verify the physical picture described above, we measured the initial charges $Q_\text{i} = 18$--90~pC  for 102 drops (variations due to stochasticity and different volumes) and tracked their evolutions $Q(t)$ over successive fission cycles $N$ (Fig.~\ref{fig:discharge}). The measurement error was $\delta Q \approx$ 0.5~pC, corresponding to $\delta Q/Q \approx 1\%$ (Materials and methods).

Before fission, drop charge is conserved, with $Q/Q_{\text{i}} = 0.97 \pm 0.02$ (dashed--dot line, Fig.~\ref{fig:discharge}a,b). Charge loss begins only after oscillations appear and fissions start ($t > 1200$ s, open markers). The temporal decay dynamics depend on the initial volume, $V = 0.5$ or \SI{1}{\micro\liter} (unfilled circles and squares, Fig.~\ref{fig:discharge}a), since larger drops require longer times to reach the critical radius $R_\text{c1}$ for the first fission.

Replotting the same $Q/Q_\text{i}$ dataset against the normalized radius $R/R_{\text{c1}}$ (Fig.~\ref{fig:discharge}b), where decreasing $R/R_{\text{c1}}$ corresponds to increasing $t$ (left to right), collapses all data onto a single universal curve $Q/Q_{\text{i}} = (R/R_{\text{c1}})^{3/2}$ as shown by the solid line. The measurements are therefore consistent with the Rayleigh limit but with a modified prefactor:  
\begin{equation} \label{eq:Q_crit_prefactor}
    Q = 4 \pi \sqrt{\gamma_{\text{eff}} \,\varepsilon_{0}}\, R_{\text{c}}^{3/2},
\end{equation}
where $\gamma_{\text{eff}} = 60$~mN\,m$^{-1}$ replaces $\gamma$ to account for the two interfaces (Fig.~\ref{fig:discharge}c). An analogous scaling holds for $R_{\text{e}}$ (for elongation) but with a slightly different prefactor, matching the small offset between between $R_{\text{e}}$ and $R_{\text{c}}$.

The dataset also follows the empirical relation $Q/Q_\text{i} = 0.98^{N}$ (Fig.~\ref{fig:discharge}d), i.e., $\Delta Q/Q \sim 2\%$ of charge loss per cycle, compared with $\sim 25\%$ for levitated drops \cite{duft_shape_2002,duft2003rayleigh,giglio2008shape, hunter2009progeny}. This striking difference explains why sessile drops sustain many more fission cycles ($N_{\text{max}} > 60$) than levitated drops ($N_{\text{max}} \approx 5$) . 

The results discussed here are true even when using a more viscous silicone oil with $\eta = \SI{100}{\milli\pascal\second}$ (green vs. blue symbols). Finally, the modified prefactor in Eq.~\ref{eq:Q_crit_prefactor} directly reflects the hemispherical geometry of sessile drops, as discussed below.

\bigskip

\begin{figure*}[!htb]
\centering
\includegraphics[scale=1]{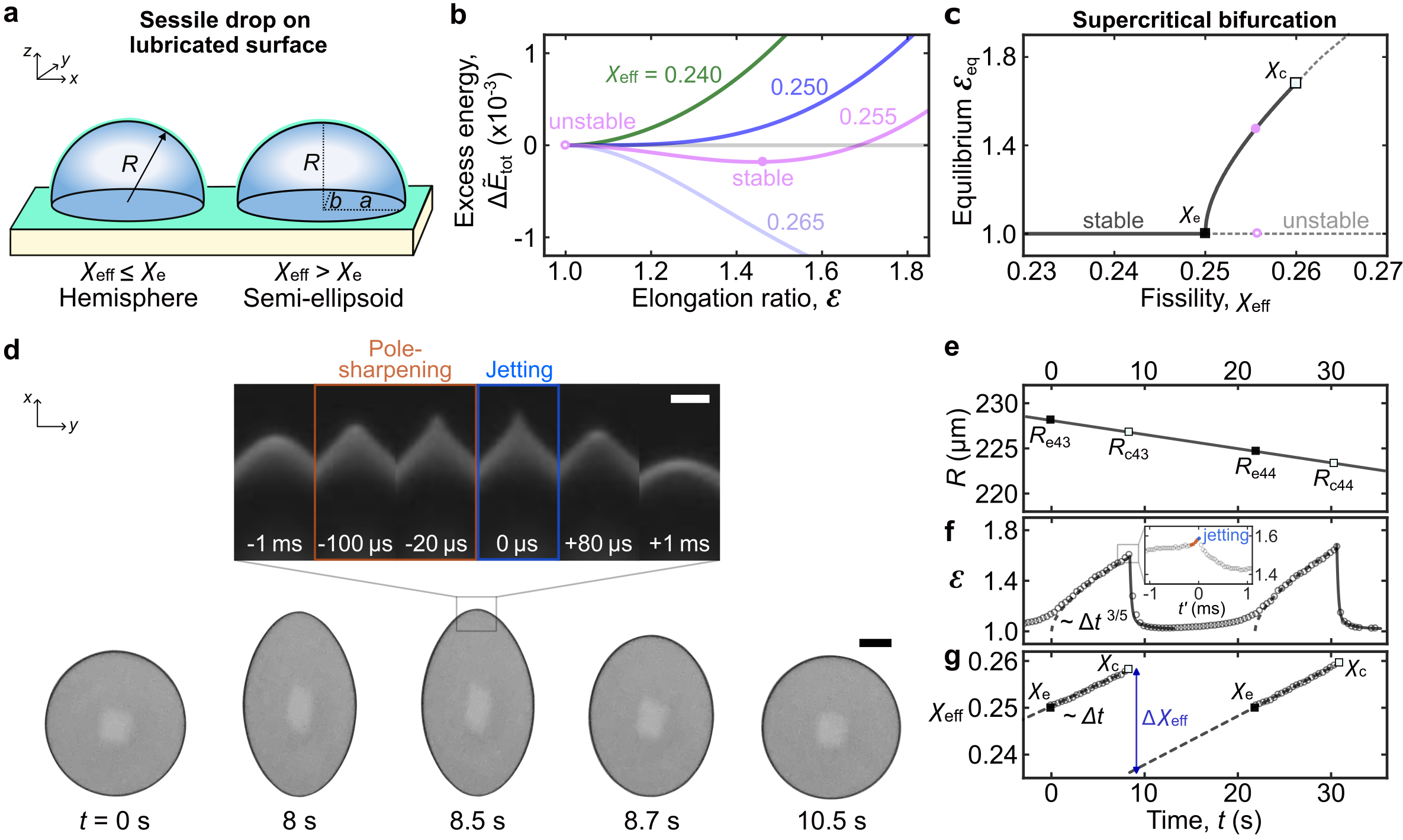}
\caption{\label{fig:bifurcation} \textbf{Fissility thresholds $X_{\text{e}}$ and $X_{\text{c}}$ for elongation and fission.} (a) Schematic of a drop elongating into a semi-ellipsoid. (b) Excess energy as a function of $\mathcal{E}$ for different $X_{\text{eff}}$ values. (c) Bifurcation diagram derived from (b), revealing a supercritical bifurcation at $X_{\text{e}} = 0.25$. Equilibrium $\mathcal{E}_{\text{eq}}$ follows the stable solution branches (solid line). Instability at $X_{\text{c}} = 0.26$ results in fission. (d) Optical snapshots over one oscillation cycle ($t$ = 0--10.5 s in panels e--g). Scale bar: \SI{100}{\micro\meter}. Inset: rapid pole-sharpening and jetting at $X_{\text{c}}$. Scale bar: \SI{50}{\micro\meter}. (e--g) Temporal evolution for $R$, $\mathcal{E}$, and $X_{\text{eff}}$ over two successive cycles $N$ = 43 and 44. Inset in (f) is derived from inset in (d), where the relative time $t' = 0$ s corresponds to onset of jetting.}
\end{figure*}

\textbf{Fissility thresholds, $X_{\text{e}}$ and $X_{\text{c}}$}. Drop elongation can be rationalised by considering the total free energy as a function of elongation ratio $\mathcal{E}$:
\begin{equation} \label{eq:E_total}
E_{\text{tot}}(\mathcal{E}) 
= 2 \pi R^{2} \gamma_{\text{eff}} \cdot f_{\text{KT}}(\mathcal{E}) 
+ \frac{1}{2} \,\frac{Q^{2}}{C(\mathcal{E})} .
\end{equation}
The first term represents the surface energy, using the Knud Thomsen approximation
\begin{equation} \label{eq:f_KT}
f_{\text{KT}}(\mathcal{E}) 
= \left[ \frac{1 + \mathcal{E}^{4/5} + \mathcal{E}^{-4/5}}{3} \right]^{5/8}
\end{equation}
to account for the surface area of a semi-ellipsoid elongating within the $x$-$y$ base plane (Fig.~\ref{fig:bifurcation}a). The second term corresponds to the electrostatic energy, with the capacitance 
\begin{equation} \label{eq:C_semi_gen}
\begin{aligned}
C(\mathcal{E}) = \frac{\alpha \cdot 4 \pi \varepsilon_{0} R}{R_{F}(\mathcal{E}, \mathcal{E}^{-1}, 1)}, 
\end{aligned}
\end{equation}
where $R_{F} (\mathcal{E}, \mathcal{E}^{-1}, 1)$ denotes Carlson's symmetric elliptic integral of the first kind, and $\alpha \approx 1/2$ is a correction factor accounting for truncation to a semi-ellipsoid \cite{shumpert1972capacitance}. $\alpha$ will shift modestly when accounting for the dielectric properties of the PMP half-space, but this secondary correction does not alter the analysis presented here.  For $\mathcal{E} = 1$ and $\alpha=1$, Eq.~\ref{eq:C_semi_gen} recovers the classical capacitance of a sphere $4 \pi \varepsilon_{0} R $.
  
The excess energy for elongation $\Delta E_{\text{tot}}(\mathcal{E}) = E_{\text{tot}}(\mathcal{E}) - E_{\text{tot}}(1)$ can be normalised by the characteristic surface energy $2 \pi R^{2} \gamma_{\text{eff}}$ to give the dimensionless form
\begin{equation} \label{eq:E_norm}
\Delta \tilde{E}_{\text{tot}}(\mathcal{E}) 
= f_{\text{KT}}(\mathcal{E}) - 1 
+ \frac{4X_{\text{eff}}}{\alpha} \,\big[ R_{F}(\mathcal{E}, \mathcal{E}^{-1}, 1) - 1 \big]. 
\end{equation}
Here,
\begin{equation} \label{eq:X_eff}
X_{\text{eff}} = \frac{Q^{2}}{64 \pi^{2} \gamma_{\text{eff}} \varepsilon_{0} R^{3} }
\end{equation}
is the effective fissility with $\gamma$ replaced by $\gamma_{\text{eff}}$ (Supplementary Fig.~\hl{S4}). 

We numerically evaluated $\Delta \tilde{E}_{\text{tot}}(\mathcal{E})$ for different $X_{\text{eff}}$ (Fig.~\ref{fig:bifurcation}b) and found that the onset of elongation occurs at the threshold value $X_{\text{e}} = 0.250$ (filled square, Fig.~\ref{fig:bifurcation}c). For $X_{\text{eff}} \leq X_{\text{e}}$, the state $\mathcal{E}_{\text{eq}} = 1$ corresponds to an energy minimum, and the hemispherical shape is stable. When $X_{\text{eff}} > X_{\text{e}}$, however, $\mathcal{E} = 1$ becomes unstable (open circle, Fig.~\ref{fig:bifurcation}b), and the energy minimum shifts continuously to $\mathcal{E}_{\text{eq}} > 1$;  for instance, at $X_{\text{eff}} = 0.255$, $\mathcal{E}_{\text{eq}} = 1.45$ (filled circle). Note that since $\Delta \tilde{E}_{\text{tot}} \sim 10^{-3}$,  even modest contact-line pinning prevents the drop from attaining $\mathcal{E}_{\text{eq}}$, which explains why spontaneous elongation and fissions are not observed on most surfaces.  

The equilibrium elongation ratio $\mathcal{E}_{\text{eq}}$ therefore follows the two solution branches (solid line in Fig.~\ref{fig:bifurcation}c):
\begin{equation} \label{eq:bifurcation}
\mathcal{E}_{\text{eq}}(X) =
\begin{cases} 
1, & X_{\text{eff}} \leq X_{\text{e}}, \\[6pt]
1 + 10.8 \,(X - X_{\text{e}})^{3/5}, & X_{\text{eff}} > X_{\text{e}},
\end{cases}
\end{equation} 
with the prefactor and exponent in the second branch obtained by fitting the numerical results with a power law. This second branch predicts a smooth increase of $\mathcal{E}_{\text{eq}} > 1$ with $X_{\text{eff}} > X_{\text{e}}$, consistent with the gradual elongation observed experimentally during the rise phase ($t$=0--8.5 s in Fig.~\ref{fig:bifurcation}d). The fact that $\mathcal{E}_{\text{eq}} > 1$ can be stable is unique to sessile drops and is characteristic of a supercritical bifurcation; levitated drops undergo a subcritical bifurcation at $ X_{\text{R}} = 1$, and no stable solution is possible for $\mathcal{E} > 1$ (Supplementary Fig.~\hl{S5, 6}). 

Our theory assumes a quasi-static equilibrium and a semi-ellipsoidal geometry throughout evaporation, leading to the unphysical prediction of indefinite elongation. In practice, elongation saturates at a peak ratio $\mathcal{E}_{\text{peak}} = 1.4$--$1.9$ set by drop size (Fig.~\ref{fig:fission}d; supplementary Fig.~\hl{S3}),  thereby defining a second scale-dependent fissility threshold $X_{\text{c}} = 0.260 \pm 0.004$ (open square, Fig.~\ref{fig:bifurcation}c). At this threshold, one pole sharpens within $\sim$1 ms forming a conical tip (Insets, Fig.~\ref{fig:bifurcation}d, f). Jetting then expels charge, driving $X_{\text{eff}}$ below $X_{\text{e}}$ and causing contraction  ($t = 8.5$--10.5 s, Fig.~\ref{fig:bifurcation}d). 

$X_{\text{c}}$ thus marks the breakdown of the semi-ellipsoidal assumption and the start of a dynamical instability not accounted for by our quasi-static theory. Importantly, the value $X_{\text{c}} = 0.26$ derived from $\mathcal{E}_{\text{peak}}$ is consistent with the modified Rayleigh limit obtained from $Q$ measurements (Eq.~\ref{eq:Q_crit_prefactor}), which predicts $(4/8)^{2} = 0.25$, underscoring the close agreement between theory and experiment. 

The bifurcation scenario can be tested rigorously against oscillation dynamics for successive cycles $N=43$ and $44$ (Fig.~\ref{fig:bifurcation}e--g). For short intervals $\Delta t$, the drop radius decreases linearly as $R(t+\Delta t) = R(t)(1 - \Delta t/\tau_{\text{evap}})$, where $\tau_{\text{evap}} = R/|\mathrm{d}R/\mathrm{d}t| = 710$~s from our experiment (Fig.~\ref{fig:fission}e). Substituting into the fissility expression and Taylor expanding about $X_{\text{e}}$ gives
\begin{equation}
\begin{aligned} \label{eq:X_Dt}
X_{\text{eff}}(\Delta t) &= \frac{Q^{2}}{64 \pi^{2} \gamma_{\text{eff}} \, \varepsilon_{0} R_{\text{e}}^{3} } \left( 1 - \frac{\Delta t}{\tau_{\text{evap}}} \right)^{-3} \\
            & \approx X_{\text{e}} \left(1 + 3 \, \frac{\Delta t}{\tau_{\text{evap}}}  \right).
\end{aligned}	
\end{equation}
Inserting this into Eq.~\ref{eq:bifurcation} (second branch) yields
\begin{equation}
\begin{aligned} \label{eq:phi_rise}
\mathcal{E}(\Delta t) &= 1 + 10.8(3X_{\text{e}})^{3/5}\left(\frac{\Delta t}{\tau_{\text{evap}}}\right)^{3/5} \\
               &= 1 + 9.09\left(\frac{\Delta t}{\tau_{\text{evap}}}\right)^{3/5},	
\end{aligned}
\end{equation}
which reproduces the rise phase of $\mathcal{E}$ without adjustable parameters, not only for $N=43$ and $44$ (dashed lines, Fig.~\ref{fig:bifurcation}f) but also for $N=30$, $52$, and $54$ (Inset 2, Fig.~\ref{fig:fission}d).

The rise time and $\Delta t_{N}$ are therefore governed by $\tau_{\text{evap}}$. In contrast, the post-fission relaxation is well captured by a sum of exponential decays \( \sum_{i=1}^{3} A_i \exp(-t/\tau_i) \) with three resolvable time constants: the fastest corresponds to the inertial-capillary time \( \tau_{1} \approx \tau_{\gamma} \), and the slowest extends to \( \sim 0.3\,\mathrm{s} \) (solid lines in Fig.~\ref{fig:bifurcation}f; supplementary Fig.~\hl{S7} and Table \hl{S1}). The asymmetry between rise and fall times give rise to the anhormic oscillations observed.

\begin{figure*}[!htb]
\centering
\includegraphics[scale=1]{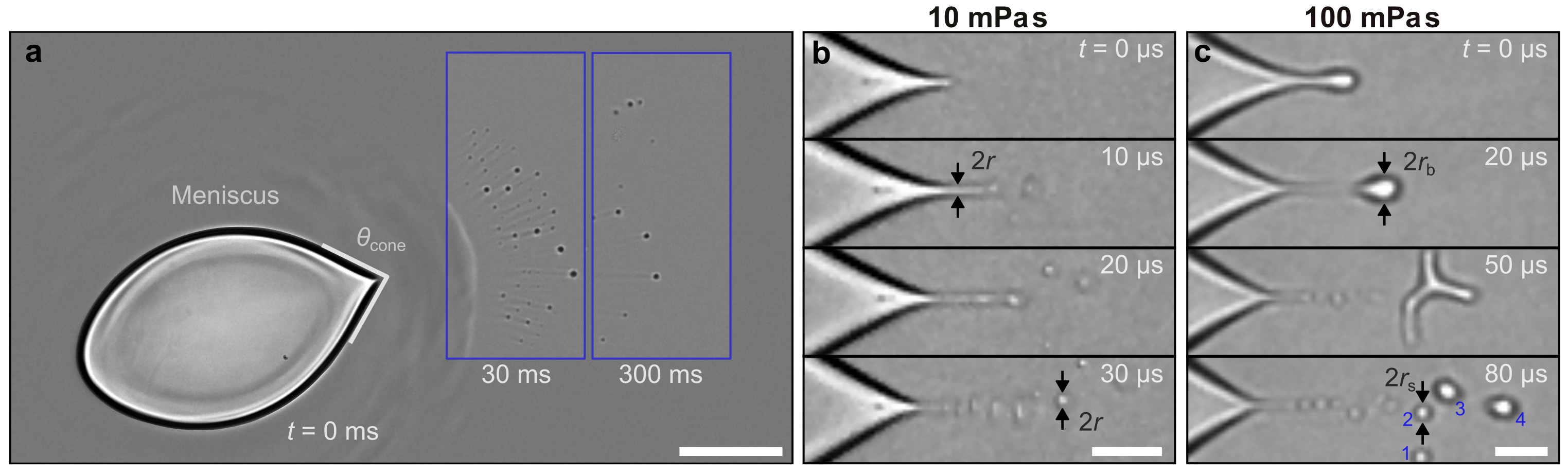}
\caption{\label{fig:jetting} \textbf{Jetting dynamics.} (a) Tear-shaped drop geometry during jetting. Insets: superimposed images of progeny microdroplets ejected at successive times. Scale bar: \SI{200}{\micro\meter}. (b) Jet formation inside the oil meniscus (viscosity \SI{10}{\milli\pascal\second}) and subsequent disintegration into progeny microdroplets, which collectively form the spray in (a). Scale bar: \SI{20}{\micro\meter}. (c) For a more viscous lubricant oil (\SI{100}{\milli\pascal\second}), a bulbous end of radius $r_{b}$  forms, which later disintegrates into four smaller secondary progeny droplets of radius $r_{s}$ (marked 1--4). Scale bar: \SI{10}{\micro\meter}.}
\end{figure*}

From Eq.~\ref{eq:bifurcation}, $X_{\text{eff}}$ is directly obtained from the measured $\mathcal{E}$ for $X_{\text{eff}} \geq X_{\text{e}}$ (markers in Fig.~\ref{fig:bifurcation}g). Backward propagation with Eq.~\ref{eq:X_Dt} then yields the temporal evolution for $X_{\text{eff}} < X_{\text{e}}$ (dashed lines, Fig.~\ref{fig:bifurcation}g). From $\Delta X_{\text{eff}} = 0.021$, we further infer a charge loss $\Delta Q/Q \approx 3\%$ for $N=43$, consistent with the direct $Q$ measurements in Fig.~\ref{fig:discharge}d.

In summary, the fissility thresholds $X_{\text{e}}$ and $X_{\text{c}}$ account for the origin of $\tau_{N}$ and of the two critical radii, $R_{\text{e}}$ and $R_{\text{c}}$ (Fig.~\ref{fig:bifurcation}e). This theoretical framework holds regardless of the lubricant viscosity $\eta = 10, \SI{100}{\milli\pascal\second}$ (Supplementary Fig.~\hl{S8}). However, the subsequent jetting dynamics differ markedly, as we discuss in the next section.

\bigskip

\textbf{Jetting and progeny microdroplets.} In sessile drops, we consistently observe one-sided jetting: a single pole sharpens into a conical tip, yielding a tear-shaped drop (Fig.~\ref{fig:jetting}a; supplementary Video~\hl{S4}). One-sided jetting has also been reported for levitated drops~\cite{gomez_charge_1994, singh2021subcritical, lazo2024self}, although the conventional view has been that of symmetric, two-sided jetting~\cite{giglio2008shape,duft2003rayleigh}. Drop deformations can be decomposed into distinct Legendre modes~\cite{rayleigh1879capillary,lamb_hydrodynamics,chang2013substrate}; the observed asymmetry in jetting suggests that the third (odd) Legendre mode also contributes to the instability, rather than the second (even) mode alone, which would favour symmetric, two-sided jetting~\cite{rayleigh_xx_1882,taylor_disintegration_1964}.

For a drop of radius $R = \SI{220}{\micro\meter}$, the measured cone angle is $\theta_{\text{cone}} = 86^{\circ}$, notably smaller than the classical Taylor cone angle of $98.6^{\circ}$ (Fig.~\ref{fig:jetting}a). This deviation reflects the dynamic, transient nature of the cone during fission, whereas Taylor's prediction applies to the static case. Moreover, we find that that cone geometry is scale dependent: $\theta_{\text{cone}}$ decreases systematically from 104$^{\circ}$ to 59$^{\circ}$ as the drop size decreases from $R$ = \SI{350}{\micro\meter} to \SI{60}{\micro\meter}, coinciding with higher $\mathcal{E}_{\text{peak}}$ for smaller drops (Supplementary Fig.~\hl{S3}).

What transpires after cone formation depends on the lubricant viscosity. For low viscosity $\eta = \SI{10}{\milli\pascal\second}$, jet formation results from a balance between capillary and electrical stresses that induce intense flow focusing at the conical tip \cite{montanero2020dripping}, producing a jet of radius $r \approx \SI{1}{\micro\meter}$ (Fig.~\ref{fig:jetting}b). The jet rapidly disintegrates within microseconds into $n = $ 40--50 progeny microdroplets that are ejected into the oil meniscus at speeds $\sim$ $\SI{1}{\meter \per \second}$ (Supplementary Video \hl{S5}). The ejected microdroplets gradually decelerate under Stokes drag, reaching speeds $\sim\SI{1}{\milli\meter\per\second}$ upon crossing the oil meniscus. They then fan out due to mutual Coulomb repulsion (Insets, Fig.~\ref{fig:jetting}a), travelling distances many times their own size. 

To assess the electrical stability of these microdroplets, each presumed to carry charge $q = \Delta Q/n$ and radius $r$, we introduce a modified fissility,
\begin{equation} \label{eq:x}
\begin{split}
    \chi &= \frac{q^{2}} {64 \pi^{2} \gamma_{\text{ow}} \, \varepsilon_{0} \varepsilon_{r}  r^{3} }, \\ 
\end{split}
\end{equation}
where $\gamma_{\text{ow}}$ replaces $\gamma_{\text{eff}}$ to reflect the oil-water interfacial tension, and $\varepsilon_{r} \approx 2.7$ is the relative permittivity of silicone oil. Dividing by $X_{\text{c}} = 0.26, $
\begin{equation} \label{eq:x_over_X}
\begin{split}
    \frac{\chi}{X_{\text{c}}} &= \left( \frac{\gamma_{\text{eff}}}{\varepsilon_{r} \gamma_{\text{ow}}} \right)  \left(\frac{q}{Q}\right)^{2} \left(\frac{R}{r}\right)^{3} \\
                   &\approx \frac{0.6}{n^{2}} \left(\frac{\Delta Q}{Q}\right)^{2} \left(\frac{R}{r}\right)^{3}, \\
\end{split}
\end{equation} 
and substituting $n$  = 45, $\Delta Q/Q = 0.02$ and $R/r = 200$ yields $\chi \approx 0.2$, indicating that the progeny microdroplets are electrically stable. Consequently, no secondary fission was observed.

In contrast, for more viscous $\eta = \SI{100}{\milli\pascal\second}$, viscous stresses become significant enough to suppress flow focusing, leading to the formation of a bulbous end of radius $r_{\text{b}} = \SI{2.5}{\micro\meter}$ with charge $q_{\text{b}}$ (Fig.~\ref{fig:jetting}c). The corresponding fissility can be written as
\begin{equation} \label{eq:x_b_over_X}
\begin{split}
    \frac{\chi_{\text{b}}}{X_{\text{c}}} &= \left( \frac{\gamma_{\text{eff}}}{\varepsilon_{r} \gamma_{\text{ow}}} \right)  \left(\frac{q_{\text{b}}}{Q}\right)^{2} \left(\frac{R}{r_{\text{b}}}\right)^{3} \\
                   &\approx 0.6 \left(\frac{R}{r_{\text{b}}}\right).\\
\end{split}
\end{equation} 
Here we take $q_{\text{b}}/Q \sim r_{\text{b}}/R$, consistent with charge equilibration and the linear scaling of capacitance with radius. (Supplementary Fig.~\hl{S9} and Table \hl{S2}). Substituting $R/r_{\text{b}} = 100$ yields $\chi_{\text{b}} \approx 15$, well above the fissility threshold. Consequently, the bulb undergoes fission within \SI{50}{\micro\second}, breaking up into four secondary progeny droplets of radius $r_{\text{s}} \approx (1/4)^{1/3} r_{\text{b}}$ (Fig.~\ref{fig:jetting}c; supplementary Video \hl{S6}). Notably, this mode of breakup differs from conventional Coulomb fission at $\chi \approx 1$, where a fine jet typically emerges.

In summary, low viscosities favour uniform, smaller progeny droplets, while higher viscosities produce larger, more polydisperse droplets through coarse breakup.
These findings highlight viscosity as a control parameter for tuning droplet size and distribution (Supplementary Videos \hl{S7, 8}), which can have important implications for nanoscale fabrication.

\bigskip
\textbf{Future outlook.} So far we have considered Coulomb fissions with no external electric field, for which the jetting direction changes between cycles. Applying a modest $E \approx 2000~\mathrm{V\,m^{-1}}$ aligns the jetting direction and thus facilitates directed collection (Supplementary Figs.~\hl{S10, 11} and Video \hl{S9}). Experimentally, the fissions are robust to added solutes, including salts and small molecules, a feature that can be harnessed for micro- and nanoscopic materials fabrication (Supplementary Figs.~\hl{S11, 12}).

The electrohydrodynamics of jet formation remains only partially understood---both in general and in the sessile drop configuration examined here \cite{montanero2020dripping, rubio2023role, collins2013universal, betelu2006singularities, ganan2016onset}. By assuming a semi-ellipsoidal geometry, our theory successfully captures the fissility threshold $X_{\text{e}}$ (for elongation) but not $X_{\text{c}}$ (for fission). A more complete theory for $X_{\text{c}}$ must relax the ellipsoidal assumption and admit solutions for conical geometries \cite{stone1999drops}. Incorporating viscous flow around the developing cone would further enable a rational prediction of progeny droplet size distributions. Finally, understanding the small charge loss $\Delta Q/Q \sim 2\%$ will require consideration of the charge-equilibration time, which is governed by the electrical conductivity of water and the dielectric properties of the surrounding oil.

\section*{Conclusion}

Our work demonstrates that modest electrical charging---arising from routine laboratory handling---can strongly influence classical drop evaporation experiments. Pipetted drops spontaneously undergo periodic Coulomb fissions governed by two distinct fissility thresholds, $X_{\text{e}} = 0.25$ for elongation and $X_{\text{c}} = 0.26$ for fission---revealing a richer instability landscape for sessile drops than for levitated ones, where a single threshold $X_{\text{R}} = 1$ controls both behaviours.

Our experimental setup is simple and robust, relying only on readily available materials---plastic Petri dishes, silicone oil, and micropipettes---to generate Coulomb fissions in millimetric-sized water drops. This work therefore provides a straightforward and accessible platform for exploring electrohydrodynamic instabilities, with potential applications in nanoscopic materials fabrication and electrospray ionization.

\section*{Materials and methods}

\textbf{Material and sample preparation.} Polymethylpentene (PMP) petri dishes were purchased from Thermo Scientific (Thermo Scientific$^{\text{TM}}$ Nalgene$^{\text{TM}}$), while silicone oils (viscosity $\eta = $ 10 and 100 mPa s$^{-1}$) were obtained from Sigma Aldrich. The DFSB-K175 fluorescent dye was purchased from Risk Reactor Inc. To prepare the lubricated surfaces, the silicone oil was spin-coated on PMP at 5000 rpm for 1 minute, resulting in a film thickness of 560 nm, as measured by white-light interferometry. 

\begin{figure}
\includegraphics[scale=0.9]{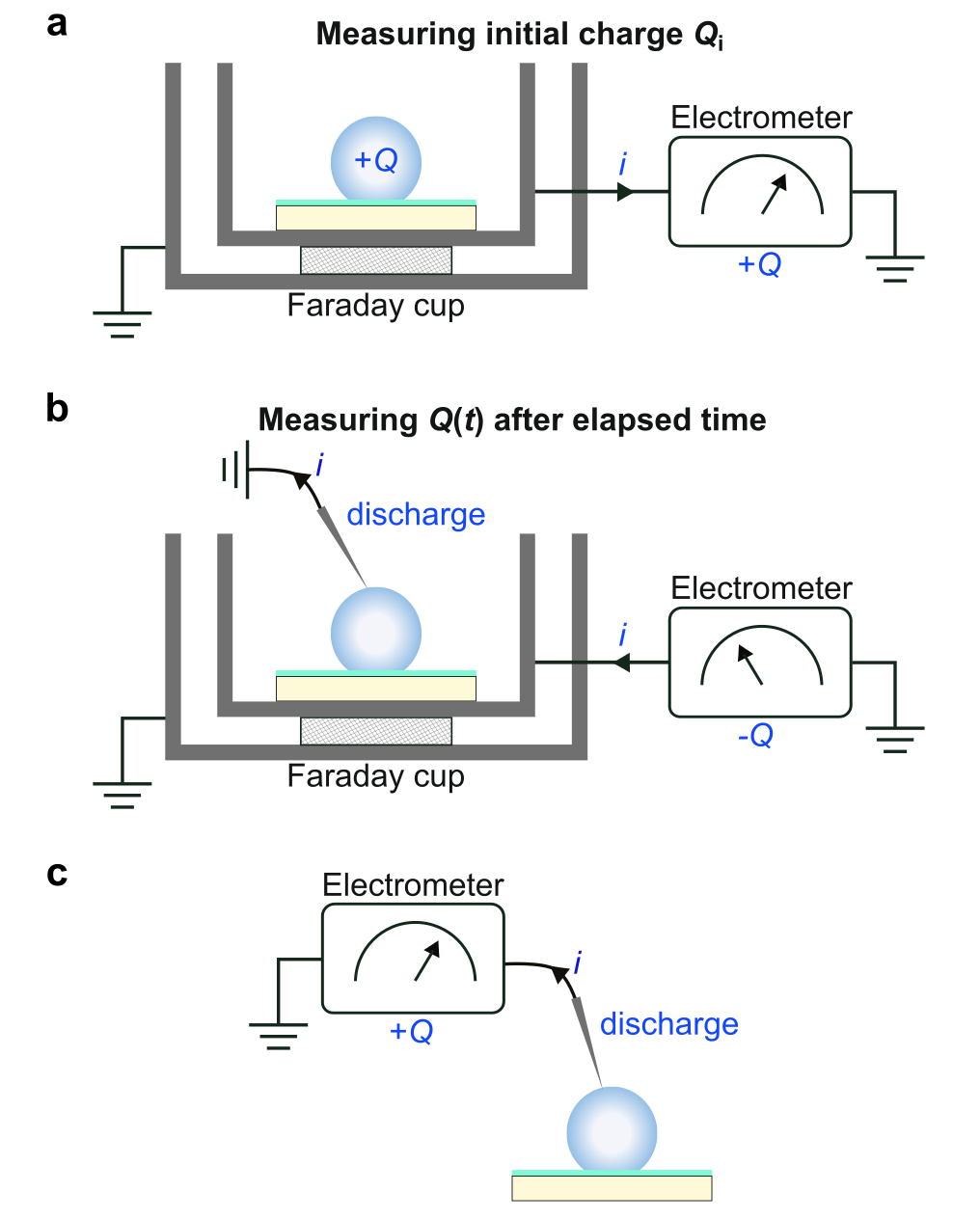}
\caption{ \textbf{Measuring drop charge $Q$.} Measuring (a) initial drop charge $Q_\text{i}$ and (b, c) the charge remaining $Q(t)$ after elapsed time.}
\label{fig:Q_measure}
\end{figure} 

\bigskip
\textbf{Surface tension measurements.} Surface and interfacial tensions were measured using the pendant drop method. To measure $\gamma_{\text{oa}}$, the silicone oil drop was suspended in air from a stainless-steel needle; for $\gamma_{\text{ow}}$, a water drop was suspended in oil. The drop profile was recorded and fitted to the Young--Laplace equation. 
\bigskip

\textbf{Visualizing meniscus skirt.} We dyed the silicone oil with 2.5\% by volume with DFSB-K175 fluorescent dye and created a fluorescent $z$-stack image using a confocal microscope (Zeiss LSM 700,  40$\times$water immersion objective). The dye was excited using a laser wavelength of 405 nm, and the fluorescent signal was collected in the range of 560-800 nm. As there is a mismatch between the refractive index of the silicone oil and the water immersion environment, the $z$-height was scaled by the factor $n_{\text{oil}}$/$n_{\text{water}}$. The profile of the wetting ridge is determined using a combination of intensity thresholding and gradient analysis, and circular fitting is used to calculate the radii $r_{\text{w}}$ and $r_{\text{o}}$.

\bigskip
\textbf{Drop deposition and charge measurement.} We deposited drops of different volumes $V = 0.5$--2 \SI{}{\micro\liter} on the surface using a conventional micropipette and polypropylene tips (Thermo Scientific Finntip). The drops naturally acquired positive charges $Q$ = 18--90 pC due to contact electrification \cite{choi2013spontaneous, wang2019origin, nauruzbayeva2020electrification}.  Previous work shows that $Q \propto V^{2/3}$, although there is significant stochasticity between drops.   

To accurately measure the initial charge $Q_\text{i}$, we deposited the drop on the lubricated PMP surface positioned inside a Faraday cup, which was connected to a Keithley 6514 electrometer (Fig.~\ref{fig:Q_measure}a; see Supplementary Fig.~\hl{S13} and Table \hl{S3} for exact dimensions). To track the remaining charge $Q$($t$) after elapsed time $t$, we discharge the drop by briefly touching it with a fine conducting platinum wire (diameter: 100 \SI{}{\micro\meter}) that is earthed. The electrometer detects a sudden decrease in electrical charge whose magnitude corresponds to that in the drop (Fig.~\ref{fig:Q_measure}b). The whole experimental setup is placed inside a Faraday cage (not shown) to reduce electronic noise. Alternatively, the drop can be discharged directly into the electrometer to measure $Q$($t$) (Fig.~\ref{fig:Q_measure}c). Both methods give similar results. 

Using our method, we measure a snapshot of $Q(t)$ in time (rather than continuously) with an accuracy of $\delta Q = 0.5$ pC. We also measure the drop radius $R(t)$ and monitor the number of fission cycles $N$ simultaneously using a top-view camera. The remaining fractional charge $Q/Q_\text{i}$ can then be obtained by dividing $Q(t)$ by $Q_\text{i}$. In other words, the results presented in Fig.~\ref{fig:discharge} are derived from snapshots of 102 individual drops, rather than continuous measurements of a single drop.

\bigskip
\textbf{Visualizing drop evaporation.} A long-distance microscope objective (Mitutoyo Plan Apo, 10$\times$ magnification) connected to a high-resolution camera (Panasonic Lumix GH5, up to 4K resolutions) is used to visualize the evaporating drop (typically at 20 fps) while inside the Faraday cup. To observe ultra-fast events (up to 300,000 fps), we used a high-speed camera (Photron FASTCAM NOVA S9). We use the circular and elliptical Hough transforms to automatically detect and measure drop radii \cite{opencv_library}. 

\section*{Data availability statement}
Data supporting the findings of this article are openly available \cite{dataset}. 

\section*{Acknowledgements}
We acknowledge insightful discussions with B.H. Tan and S. Hardt. We are also indebted to J.W. Strutt for timeless inspiration.
\bigskip
\section*{Conflicts of interest}
There are no conflicts to declare.

\section*{Author Contributions}
M.L. performed the majority of the experiments, while P.Z. obtained the primary dataset in Fig. 1. O.L. and S.A. contributed to additional experiments. A.D.R. and D.D. jointly developed the theoretical framework. M.L., A.D.R., and D.D. analysed the data. D.D. conceptualized the project, supervised the research, and wrote the manuscript. All authors reviewed and approved the final manuscript.

% \clearpage
% \bibliography{main}

\begin{thebibliography}{62}%
\makeatletter
\providecommand \@ifxundefined [1]{%
 \@ifx{#1\undefined}
}%
\providecommand \@ifnum [1]{%
 \ifnum #1\expandafter \@firstoftwo
 \else \expandafter \@secondoftwo
 \fi
}%
\providecommand \@ifx [1]{%
 \ifx #1\expandafter \@firstoftwo
 \else \expandafter \@secondoftwo
 \fi
}%
\providecommand \natexlab [1]{#1}%
\providecommand \enquote  [1]{``#1''}%
\providecommand \bibnamefont  [1]{#1}%
\providecommand \bibfnamefont [1]{#1}%
\providecommand \citenamefont [1]{#1}%
\providecommand \href@noop [0]{\@secondoftwo}%
\providecommand \href [0]{\begingroup \@sanitize@url \@href}%
\providecommand \@href[1]{\@@startlink{#1}\@@href}%
\providecommand \@@href[1]{\endgroup#1\@@endlink}%
\providecommand \@sanitize@url [0]{\catcode `\\12\catcode `\$12\catcode `\&12\catcode `\#12\catcode `\^12\catcode `\_12\catcode `\%12\relax}%
\providecommand \@@startlink[1]{}%
\providecommand \@@endlink[0]{}%
\providecommand \url  [0]{\begingroup\@sanitize@url \@url }%
\providecommand \@url [1]{\endgroup\@href {#1}{\urlprefix }}%
\providecommand \urlprefix  [0]{URL }%
\providecommand \Eprint [0]{\href }%
\providecommand \doibase [0]{http://dx.doi.org/}%
\providecommand \selectlanguage [0]{\@gobble}%
\providecommand \bibinfo  [0]{\@secondoftwo}%
\providecommand \bibfield  [0]{\@secondoftwo}%
\providecommand \translation [1]{[#1]}%
\providecommand \BibitemOpen [0]{}%
\providecommand \bibitemStop [0]{}%
\providecommand \bibitemNoStop [0]{.\EOS\space}%
\providecommand \EOS [0]{\spacefactor3000\relax}%
\providecommand \BibitemShut  [1]{\csname bibitem#1\endcsname}%
\let\auto@bib@innerbib\@empty
%</preamble>
\bibitem [{\citenamefont {Gaskell}\ \emph {et~al.}(1977)\citenamefont {Gaskell}, \citenamefont {Illingworth}, \citenamefont {Latham},\ and\ \citenamefont {Moore}}]{gaskell1977airborne}%
  \BibitemOpen
  \bibfield  {author} {\bibinfo {author} {\bibfnamefont {W.}~\bibnamefont {Gaskell}}, \bibinfo {author} {\bibfnamefont {A.~J.}\ \bibnamefont {Illingworth}}, \bibinfo {author} {\bibfnamefont {J.}~\bibnamefont {Latham}}, \ and\ \bibinfo {author} {\bibfnamefont {C.~B.}\ \bibnamefont {Moore}},\ }\bibfield  {title} {\enquote {\bibinfo {title} {Airborne studies of thunderstorm electrification},}\ }\href@noop {} {\bibfield  {journal} {\bibinfo  {journal} {Nature}\ }\textbf {\bibinfo {volume} {268}},\ \bibinfo {pages} {124--125} (\bibinfo {year} {1977})}\BibitemShut {NoStop}%
\bibitem [{\citenamefont {Feynman}\ \emph {et~al.}(1964)\citenamefont {Feynman}, \citenamefont {Leighton},\ and\ \citenamefont {Sands}}]{feynman1964electricity}%
  \BibitemOpen
  \bibfield  {author} {\bibinfo {author} {\bibfnamefont {R.~P.}\ \bibnamefont {Feynman}}, \bibinfo {author} {\bibfnamefont {R.~B.}\ \bibnamefont {Leighton}}, \ and\ \bibinfo {author} {\bibfnamefont {M.L.}\ \bibnamefont {Sands}},\ }\bibfield  {title} {\enquote {\bibinfo {title} {Electricity in the atmosphere},}\ }\href@noop {} {\bibfield  {journal} {\bibinfo  {journal} {The Feynman Lectures on Physics}\ }\textbf {\bibinfo {volume} {2}} (\bibinfo {year} {1964})}\BibitemShut {NoStop}%
\bibitem [{\citenamefont {Park}\ \emph {et~al.}(2007)\citenamefont {Park}, \citenamefont {Hardy}, \citenamefont {Kang}, \citenamefont {Barton}, \citenamefont {Adair}, \citenamefont {Mukhopadhyay}, \citenamefont {Lee}, \citenamefont {Strano}, \citenamefont {Alleyne}, \citenamefont {Georgiadis}, \citenamefont {Ferreira},\ and\ \citenamefont {Rogers}}]{park_high-resolution_2007}%
  \BibitemOpen
  \bibfield  {author} {\bibinfo {author} {\bibfnamefont {J.-U.}\ \bibnamefont {Park}}, \bibinfo {author} {\bibfnamefont {M.}~\bibnamefont {Hardy}}, \bibinfo {author} {\bibfnamefont {S.~J.}\ \bibnamefont {Kang}}, \bibinfo {author} {\bibfnamefont {K.}~\bibnamefont {Barton}}, \bibinfo {author} {\bibfnamefont {K.}~\bibnamefont {Adair}}, \bibinfo {author} {\bibfnamefont {D.~k.}\ \bibnamefont {Mukhopadhyay}}, \bibinfo {author} {\bibfnamefont {C.~Y.}\ \bibnamefont {Lee}}, \bibinfo {author} {\bibfnamefont {M.~S.}\ \bibnamefont {Strano}}, \bibinfo {author} {\bibfnamefont {A.~G.}\ \bibnamefont {Alleyne}}, \bibinfo {author} {\bibfnamefont {J.~G.}\ \bibnamefont {Georgiadis}}, \bibinfo {author} {\bibfnamefont {P.~M.}\ \bibnamefont {Ferreira}}, \ and\ \bibinfo {author} {\bibfnamefont {J.~A.}\ \bibnamefont {Rogers}},\ }\bibfield  {title} {\enquote {\bibinfo {title} {High-resolution electrohydrodynamic jet printing},}\ }\href {\doibase 10.1038/nmat1974} {\bibfield  {journal} {\bibinfo  {journal} {Nat. Mater.}\ }\textbf
  {\bibinfo {volume} {6}},\ \bibinfo {pages} {782--789} (\bibinfo {year} {2007})}\BibitemShut {NoStop}%
\bibitem [{\citenamefont {Gomez}\ and\ \citenamefont {Tang}(1994)}]{gomez_charge_1994}%
  \BibitemOpen
  \bibfield  {author} {\bibinfo {author} {\bibfnamefont {A.}~\bibnamefont {Gomez}}\ and\ \bibinfo {author} {\bibfnamefont {K.}~\bibnamefont {Tang}},\ }\bibfield  {title} {\enquote {\bibinfo {title} {Charge and fission of droplets in electrostatic sprays},}\ }\href {\doibase 10.1063/1.868037} {\bibfield  {journal} {\bibinfo  {journal} {Phys. Fluids}\ }\textbf {\bibinfo {volume} {6}},\ \bibinfo {pages} {404--414} (\bibinfo {year} {1994})}\BibitemShut {NoStop}%
\bibitem [{\citenamefont {Fenn}\ \emph {et~al.}(1989)\citenamefont {Fenn}, \citenamefont {Mann}, \citenamefont {Meng}, \citenamefont {Wong},\ and\ \citenamefont {Whitehouse}}]{fenn1989electrospray}%
  \BibitemOpen
  \bibfield  {author} {\bibinfo {author} {\bibfnamefont {J.~B.}\ \bibnamefont {Fenn}}, \bibinfo {author} {\bibfnamefont {M.}~\bibnamefont {Mann}}, \bibinfo {author} {\bibfnamefont {C.~K.}\ \bibnamefont {Meng}}, \bibinfo {author} {\bibfnamefont {S.~F.}\ \bibnamefont {Wong}}, \ and\ \bibinfo {author} {\bibfnamefont {C.~M.}\ \bibnamefont {Whitehouse}},\ }\bibfield  {title} {\enquote {\bibinfo {title} {Electrospray ionization for mass spectrometry of large biomolecules},}\ }\href@noop {} {\bibfield  {journal} {\bibinfo  {journal} {Science}\ }\textbf {\bibinfo {volume} {246}},\ \bibinfo {pages} {64--71} (\bibinfo {year} {1989})}\BibitemShut {NoStop}%
\bibitem [{\citenamefont {Rayleigh}(1882)}]{rayleigh_xx_1882}%
  \BibitemOpen
  \bibfield  {author} {\bibinfo {author} {\bibfnamefont {Lord}\ \bibnamefont {Rayleigh}},\ }\bibfield  {title} {\enquote {\bibinfo {title} {On the equilibrium of liquid conducting masses charged with electricity},}\ }\href {\doibase 10.1080/14786448208628425} {\bibfield  {journal} {\bibinfo  {journal} {Philos. Mag.}\ }\textbf {\bibinfo {volume} {14}},\ \bibinfo {pages} {184--186} (\bibinfo {year} {1882})}\BibitemShut {NoStop}%
\bibitem [{\citenamefont {Duft}\ \emph {et~al.}(2003)\citenamefont {Duft}, \citenamefont {Achtzehn}, \citenamefont {M{\"u}ller}, \citenamefont {Huber},\ and\ \citenamefont {Leisner}}]{duft2003rayleigh}%
  \BibitemOpen
  \bibfield  {author} {\bibinfo {author} {\bibfnamefont {D.}~\bibnamefont {Duft}}, \bibinfo {author} {\bibfnamefont {T.}~\bibnamefont {Achtzehn}}, \bibinfo {author} {\bibfnamefont {R.}~\bibnamefont {M{\"u}ller}}, \bibinfo {author} {\bibfnamefont {B.~A.}\ \bibnamefont {Huber}}, \ and\ \bibinfo {author} {\bibfnamefont {T.}~\bibnamefont {Leisner}},\ }\bibfield  {title} {\enquote {\bibinfo {title} {Rayleigh jets from levitated microdroplets},}\ }\href@noop {} {\bibfield  {journal} {\bibinfo  {journal} {Nature}\ }\textbf {\bibinfo {volume} {421}},\ \bibinfo {pages} {128--128} (\bibinfo {year} {2003})}\BibitemShut {NoStop}%
\bibitem [{\citenamefont {Li}\ \emph {et~al.}(2005)\citenamefont {Li}, \citenamefont {Tu},\ and\ \citenamefont {Ray}}]{li_charge_2005}%
  \BibitemOpen
  \bibfield  {author} {\bibinfo {author} {\bibfnamefont {K.-Y.}\ \bibnamefont {Li}}, \bibinfo {author} {\bibfnamefont {H.}~\bibnamefont {Tu}}, \ and\ \bibinfo {author} {\bibfnamefont {A.~K.}\ \bibnamefont {Ray}},\ }\bibfield  {title} {\enquote {\bibinfo {title} {Charge limits on droplets during evaporation},}\ }\href {\doibase 10.1021/la047973n} {\bibfield  {journal} {\bibinfo  {journal} {Langmuir}\ }\textbf {\bibinfo {volume} {21}},\ \bibinfo {pages} {3786--3794} (\bibinfo {year} {2005})}\BibitemShut {NoStop}%
\bibitem [{\citenamefont {Doyle}\ \emph {et~al.}(1964)\citenamefont {Doyle}, \citenamefont {Moffett},\ and\ \citenamefont {Vonnegut}}]{doyle_behavior_1964}%
  \BibitemOpen
  \bibfield  {author} {\bibinfo {author} {\bibfnamefont {A.}~\bibnamefont {Doyle}}, \bibinfo {author} {\bibfnamefont {D.~R.}\ \bibnamefont {Moffett}}, \ and\ \bibinfo {author} {\bibfnamefont {B.}~\bibnamefont {Vonnegut}},\ }\bibfield  {title} {\enquote {\bibinfo {title} {Behavior of evaporating electrically charged droplets},}\ }\href {\doibase 10.1016/0095-8522(64)90024-8} {\bibfield  {journal} {\bibinfo  {journal} {J. Colloid Sci.}\ }\textbf {\bibinfo {volume} {19}},\ \bibinfo {pages} {136--143} (\bibinfo {year} {1964})}\BibitemShut {NoStop}%
\bibitem [{\citenamefont {Abbas}\ and\ \citenamefont {Latham}(1967)}]{abbas_instability_1967}%
  \BibitemOpen
  \bibfield  {author} {\bibinfo {author} {\bibfnamefont {M.~A.}\ \bibnamefont {Abbas}}\ and\ \bibinfo {author} {\bibfnamefont {J.}~\bibnamefont {Latham}},\ }\bibfield  {title} {\enquote {\bibinfo {title} {The instability of evaporating charged drops},}\ }\href {\doibase 10.1017/S0022112067001685} {\bibfield  {journal} {\bibinfo  {journal} {J. Fluid Mech.}\ }\textbf {\bibinfo {volume} {30}},\ \bibinfo {pages} {663--670} (\bibinfo {year} {1967})}\BibitemShut {NoStop}%
\bibitem [{\citenamefont {Taylor}(1964)}]{taylor_disintegration_1964}%
  \BibitemOpen
  \bibfield  {author} {\bibinfo {author} {\bibfnamefont {G.~I.}\ \bibnamefont {Taylor}},\ }\bibfield  {title} {\enquote {\bibinfo {title} {Disintegration of water drops in an electric field},}\ }\href {\doibase 10.1098/rspa.1964.0151} {\bibfield  {journal} {\bibinfo  {journal} {Proc. R. Soc. Lond. A}\ }\textbf {\bibinfo {volume} {280}},\ \bibinfo {pages} {383--397} (\bibinfo {year} {1964})}\BibitemShut {NoStop}%
\bibitem [{\citenamefont {Basaran}\ and\ \citenamefont {Scriven}(1989)}]{basaran_axisymmetric_1989}%
  \BibitemOpen
  \bibfield  {author} {\bibinfo {author} {\bibfnamefont {O.~A.}\ \bibnamefont {Basaran}}\ and\ \bibinfo {author} {\bibfnamefont {L.~E.}\ \bibnamefont {Scriven}},\ }\bibfield  {title} {\enquote {\bibinfo {title} {Axisymmetric shapes and stability of charged drops in an external electric field},}\ }\href {\doibase 10.1063/1.857377} {\bibfield  {journal} {\bibinfo  {journal} {Phys. Fluids A}\ }\textbf {\bibinfo {volume} {1}},\ \bibinfo {pages} {799--809} (\bibinfo {year} {1989})}\BibitemShut {NoStop}%
\bibitem [{\citenamefont {Fernández de~la Mora}(1996)}]{fernandez_de_la_mora_outcome_1996}%
  \BibitemOpen
  \bibfield  {author} {\bibinfo {author} {\bibfnamefont {J.}~\bibnamefont {Fernández de~la Mora}},\ }\bibfield  {title} {\enquote {\bibinfo {title} {On the outcome of the {Coulombic} fission of a charged isolated drop},}\ }\href {\doibase 10.1006/jcis.1996.0109} {\bibfield  {journal} {\bibinfo  {journal} {J. Colloid Interface Sci.}\ }\textbf {\bibinfo {volume} {178}},\ \bibinfo {pages} {209--218} (\bibinfo {year} {1996})}\BibitemShut {NoStop}%
\bibitem [{\citenamefont {Achtzehn}\ \emph {et~al.}(2005)\citenamefont {Achtzehn}, \citenamefont {Müller}, \citenamefont {Duft},\ and\ \citenamefont {Leisner}}]{achtzehn_coulomb_2005}%
  \BibitemOpen
  \bibfield  {author} {\bibinfo {author} {\bibfnamefont {T.}~\bibnamefont {Achtzehn}}, \bibinfo {author} {\bibfnamefont {R.}~\bibnamefont {Müller}}, \bibinfo {author} {\bibfnamefont {D.}~\bibnamefont {Duft}}, \ and\ \bibinfo {author} {\bibfnamefont {T.}~\bibnamefont {Leisner}},\ }\bibfield  {title} {\enquote {\bibinfo {title} {The {Coulomb} instability of charged microdroplets: dynamics and scaling},}\ }\href {\doibase 10.1140/epjd/e2005-00102-1} {\bibfield  {journal} {\bibinfo  {journal} {Eur. Phys. J. D}\ }\textbf {\bibinfo {volume} {34}},\ \bibinfo {pages} {311--313} (\bibinfo {year} {2005})}\BibitemShut {NoStop}%
\bibitem [{\citenamefont {Taflin}\ \emph {et~al.}(1989)\citenamefont {Taflin}, \citenamefont {Ward},\ and\ \citenamefont {Davis}}]{taflin_electrified_1989}%
  \BibitemOpen
  \bibfield  {author} {\bibinfo {author} {\bibfnamefont {D.~C.}\ \bibnamefont {Taflin}}, \bibinfo {author} {\bibfnamefont {T.~L.}\ \bibnamefont {Ward}}, \ and\ \bibinfo {author} {\bibfnamefont {E.~J.}\ \bibnamefont {Davis}},\ }\bibfield  {title} {\enquote {\bibinfo {title} {{Electrified droplet fission and the Rayleigh limit}},}\ }\href {\doibase 10.1021/la00086a016} {\bibfield  {journal} {\bibinfo  {journal} {Langmuir}\ }\textbf {\bibinfo {volume} {5}},\ \bibinfo {pages} {376--384} (\bibinfo {year} {1989})}\BibitemShut {NoStop}%
\bibitem [{\citenamefont {Davis}\ and\ \citenamefont {Bridges}(1994)}]{davis_rayleigh_1994}%
  \BibitemOpen
  \bibfield  {author} {\bibinfo {author} {\bibfnamefont {E.~J.}\ \bibnamefont {Davis}}\ and\ \bibinfo {author} {\bibfnamefont {M.~A.}\ \bibnamefont {Bridges}},\ }\bibfield  {title} {\enquote {\bibinfo {title} {{The Rayleigh limit of charge revisited: light scattering from exploding droplets}},}\ }\href {\doibase 10.1016/0021-8502(94)90208-9} {\bibfield  {journal} {\bibinfo  {journal} {J. Aerosol Sci.}\ }\textbf {\bibinfo {volume} {25}},\ \bibinfo {pages} {1179--1199} (\bibinfo {year} {1994})}\BibitemShut {NoStop}%
\bibitem [{\citenamefont {Grimm}\ and\ \citenamefont {Beauchamp}(2005)}]{grimm_dynamics_2005}%
  \BibitemOpen
  \bibfield  {author} {\bibinfo {author} {\bibfnamefont {R.~L.}\ \bibnamefont {Grimm}}\ and\ \bibinfo {author} {\bibfnamefont {J.~L.}\ \bibnamefont {Beauchamp}},\ }\bibfield  {title} {\enquote {\bibinfo {title} {Dynamics of field-induced droplet ionization: time-resolved studies of distortion, jetting, and progeny formation from charged and neutral methanol droplets exposed to strong electric fields},}\ }\href {\doibase 10.1021/jp0450540} {\bibfield  {journal} {\bibinfo  {journal} {J. Phys. Chem. B}\ }\textbf {\bibinfo {volume} {109}},\ \bibinfo {pages} {8244--8250} (\bibinfo {year} {2005})}\BibitemShut {NoStop}%
\bibitem [{\citenamefont {Duft}\ \emph {et~al.}(2002)\citenamefont {Duft}, \citenamefont {Lebius}, \citenamefont {Huber}, \citenamefont {Guet},\ and\ \citenamefont {Leisner}}]{duft_shape_2002}%
  \BibitemOpen
  \bibfield  {author} {\bibinfo {author} {\bibfnamefont {D.}~\bibnamefont {Duft}}, \bibinfo {author} {\bibfnamefont {H.}~\bibnamefont {Lebius}}, \bibinfo {author} {\bibfnamefont {B.~A.}\ \bibnamefont {Huber}}, \bibinfo {author} {\bibfnamefont {C.}~\bibnamefont {Guet}}, \ and\ \bibinfo {author} {\bibfnamefont {T.}~\bibnamefont {Leisner}},\ }\bibfield  {title} {\enquote {\bibinfo {title} {Shape oscillations and stability of charged microdroplets},}\ }\href {\doibase 10.1103/PhysRevLett.89.084503} {\bibfield  {journal} {\bibinfo  {journal} {Phys. Rev. Lett.}\ }\textbf {\bibinfo {volume} {89}},\ \bibinfo {pages} {084503} (\bibinfo {year} {2002})}\BibitemShut {NoStop}%
\bibitem [{\citenamefont {Grimm}\ and\ \citenamefont {Beauchamp}(2003)}]{grimm_field-induced_2003}%
  \BibitemOpen
  \bibfield  {author} {\bibinfo {author} {\bibfnamefont {R.~L.}\ \bibnamefont {Grimm}}\ and\ \bibinfo {author} {\bibfnamefont {J.~L.}\ \bibnamefont {Beauchamp}},\ }\bibfield  {title} {\enquote {\bibinfo {title} {Field-induced droplet ionization mass spectrometry},}\ }\href {\doibase 10.1021/jp037099r} {\bibfield  {journal} {\bibinfo  {journal} {J. Phys. Chem. B}\ }\textbf {\bibinfo {volume} {107}},\ \bibinfo {pages} {14161--14163} (\bibinfo {year} {2003})}\BibitemShut {NoStop}%
\bibitem [{\citenamefont {Singh}\ \emph {et~al.}(2021{\natexlab{a}})\citenamefont {Singh}, \citenamefont {Gawande}, \citenamefont {Mayya},\ and\ \citenamefont {Thaokar}}]{singh_subcritical_2021}%
  \BibitemOpen
  \bibfield  {author} {\bibinfo {author} {\bibfnamefont {M.}~\bibnamefont {Singh}}, \bibinfo {author} {\bibfnamefont {N.}~\bibnamefont {Gawande}}, \bibinfo {author} {\bibfnamefont {Y.~S.}\ \bibnamefont {Mayya}}, \ and\ \bibinfo {author} {\bibfnamefont {R.}~\bibnamefont {Thaokar}},\ }\bibfield  {title} {\enquote {\bibinfo {title} {Subcritical asymmetric {Rayleigh} breakup of a charged drop induced by finite amplitude perturbations in a quadrupole trap},}\ }\href {\doibase 10.1103/PhysRevE.103.053111} {\bibfield  {journal} {\bibinfo  {journal} {Phys. Rev. E}\ }\textbf {\bibinfo {volume} {103}},\ \bibinfo {pages} {053111} (\bibinfo {year} {2021}{\natexlab{a}})}\BibitemShut {NoStop}%
\bibitem [{\citenamefont {Wilson}\ and\ \citenamefont {D'Ambrosio}(2023)}]{wilson2023evaporation}%
  \BibitemOpen
  \bibfield  {author} {\bibinfo {author} {\bibfnamefont {S.~K.}\ \bibnamefont {Wilson}}\ and\ \bibinfo {author} {\bibfnamefont {H.-M.}\ \bibnamefont {D'Ambrosio}},\ }\bibfield  {title} {\enquote {\bibinfo {title} {Evaporation of sessile droplets},}\ }\href@noop {} {\bibfield  {journal} {\bibinfo  {journal} {Ann. Rev. Fluid Mech.}\ }\textbf {\bibinfo {volume} {55}},\ \bibinfo {pages} {481--509} (\bibinfo {year} {2023})}\BibitemShut {NoStop}%
\bibitem [{\citenamefont {Chen}\ \emph {et~al.}(2012)\citenamefont {Chen}, \citenamefont {Ma}, \citenamefont {Li}, \citenamefont {Hao}, \citenamefont {Guo}, \citenamefont {Luk}, \citenamefont {Li}, \citenamefont {Yao},\ and\ \citenamefont {Wang}}]{chen_evaporation_2012}%
  \BibitemOpen
  \bibfield  {author} {\bibinfo {author} {\bibfnamefont {X.}~\bibnamefont {Chen}}, \bibinfo {author} {\bibfnamefont {R.}~\bibnamefont {Ma}}, \bibinfo {author} {\bibfnamefont {J.}~\bibnamefont {Li}}, \bibinfo {author} {\bibfnamefont {C.}~\bibnamefont {Hao}}, \bibinfo {author} {\bibfnamefont {W.}~\bibnamefont {Guo}}, \bibinfo {author} {\bibfnamefont {B.~L.}\ \bibnamefont {Luk}}, \bibinfo {author} {\bibfnamefont {S.~C.}\ \bibnamefont {Li}}, \bibinfo {author} {\bibfnamefont {S.}~\bibnamefont {Yao}}, \ and\ \bibinfo {author} {\bibfnamefont {Z.}~\bibnamefont {Wang}},\ }\bibfield  {title} {\enquote {\bibinfo {title} {Evaporation of droplets on superhydrophobic surfaces: {Surface} roughness and small droplet size effects},}\ }\href {\doibase 10.1103/PhysRevLett.109.116101} {\bibfield  {journal} {\bibinfo  {journal} {Phys. Rev. Lett.}\ }\textbf {\bibinfo {volume} {109}},\ \bibinfo {pages} {116101} (\bibinfo {year} {2012})}\BibitemShut {NoStop}%
\bibitem [{\citenamefont {Guan}\ \emph {et~al.}(2015{\natexlab{a}})\citenamefont {Guan}, \citenamefont {Wells}, \citenamefont {Xu}, \citenamefont {McHale}, \citenamefont {Wood}, \citenamefont {Martin},\ and\ \citenamefont {Stuart-Cole}}]{guan_evaporation_2015}%
  \BibitemOpen
  \bibfield  {author} {\bibinfo {author} {\bibfnamefont {J.~H.}\ \bibnamefont {Guan}}, \bibinfo {author} {\bibfnamefont {G.~G.}\ \bibnamefont {Wells}}, \bibinfo {author} {\bibfnamefont {B.}~\bibnamefont {Xu}}, \bibinfo {author} {\bibfnamefont {G.}~\bibnamefont {McHale}}, \bibinfo {author} {\bibfnamefont {D.}~\bibnamefont {Wood}}, \bibinfo {author} {\bibfnamefont {J.}~\bibnamefont {Martin}}, \ and\ \bibinfo {author} {\bibfnamefont {S.}~\bibnamefont {Stuart-Cole}},\ }\bibfield  {title} {\enquote {\bibinfo {title} {Evaporation of sessile droplets on slippery liquid-infused porous surfaces ({SLIPS})},}\ }\href {\doibase 10.1021/acs.langmuir.5b03240} {\bibfield  {journal} {\bibinfo  {journal} {Langmuir}\ }\textbf {\bibinfo {volume} {31}},\ \bibinfo {pages} {11781--11789} (\bibinfo {year} {2015}{\natexlab{a}})}\BibitemShut {NoStop}%
\bibitem [{\citenamefont {Charitatos}\ and\ \citenamefont {Kumar}(2021)}]{charitatos_droplet_2021}%
  \BibitemOpen
  \bibfield  {author} {\bibinfo {author} {\bibfnamefont {V.}~\bibnamefont {Charitatos}}\ and\ \bibinfo {author} {\bibfnamefont {S.}~\bibnamefont {Kumar}},\ }\bibfield  {title} {\enquote {\bibinfo {title} {Droplet evaporation on soft solid substrates},}\ }\href {\doibase 10.1039/D1SM00828E} {\bibfield  {journal} {\bibinfo  {journal} {Soft Matter}\ }\textbf {\bibinfo {volume} {17}},\ \bibinfo {pages} {9339--9352} (\bibinfo {year} {2021})}\BibitemShut {NoStop}%
\bibitem [{\citenamefont {Erbil}\ and\ \citenamefont {McHale}(2023)}]{erbil_droplet_2023}%
  \BibitemOpen
  \bibfield  {author} {\bibinfo {author} {\bibfnamefont {H.~Y.}\ \bibnamefont {Erbil}}\ and\ \bibinfo {author} {\bibfnamefont {G.}~\bibnamefont {McHale}},\ }\bibfield  {title} {\enquote {\bibinfo {title} {Droplet evaporation on superhydrophobic surfaces},}\ }\href {\doibase 10.1063/5.0159112} {\bibfield  {journal} {\bibinfo  {journal} {Appl. Phys. Lett.}\ }\textbf {\bibinfo {volume} {123}},\ \bibinfo {pages} {080501} (\bibinfo {year} {2023})}\BibitemShut {NoStop}%
\bibitem [{\citenamefont {Choi}\ \emph {et~al.}(2013)\citenamefont {Choi}, \citenamefont {Lee}, \citenamefont {Im}, \citenamefont {Kang}, \citenamefont {Lim}, \citenamefont {Kim},\ and\ \citenamefont {Kang}}]{choi2013spontaneous}%
  \BibitemOpen
  \bibfield  {author} {\bibinfo {author} {\bibfnamefont {D.}~\bibnamefont {Choi}}, \bibinfo {author} {\bibfnamefont {H.}~\bibnamefont {Lee}}, \bibinfo {author} {\bibfnamefont {D.~J.}\ \bibnamefont {Im}}, \bibinfo {author} {\bibfnamefont {I.~S.}\ \bibnamefont {Kang}}, \bibinfo {author} {\bibfnamefont {G.}~\bibnamefont {Lim}}, \bibinfo {author} {\bibfnamefont {D.~S.}\ \bibnamefont {Kim}}, \ and\ \bibinfo {author} {\bibfnamefont {K.~H.}\ \bibnamefont {Kang}},\ }\bibfield  {title} {\enquote {\bibinfo {title} {Spontaneous electrical charging of droplets by conventional pipetting},}\ }\href@noop {} {\bibfield  {journal} {\bibinfo  {journal} {Sci. Rep.}\ }\textbf {\bibinfo {volume} {3}},\ \bibinfo {pages} {2037} (\bibinfo {year} {2013})}\BibitemShut {NoStop}%
\bibitem [{\citenamefont {Wang}\ and\ \citenamefont {Wang}(2019)}]{wang2019origin}%
  \BibitemOpen
  \bibfield  {author} {\bibinfo {author} {\bibfnamefont {Z.~L.}\ \bibnamefont {Wang}}\ and\ \bibinfo {author} {\bibfnamefont {A.~C.}\ \bibnamefont {Wang}},\ }\bibfield  {title} {\enquote {\bibinfo {title} {On the origin of contact-electrification},}\ }\href@noop {} {\bibfield  {journal} {\bibinfo  {journal} {Mater. Today}\ }\textbf {\bibinfo {volume} {30}},\ \bibinfo {pages} {34--51} (\bibinfo {year} {2019})}\BibitemShut {NoStop}%
\bibitem [{\citenamefont {Nauruzbayeva}\ \emph {et~al.}(2020)\citenamefont {Nauruzbayeva}, \citenamefont {Sun}, \citenamefont {Gallo~Jr}, \citenamefont {Ibrahim}, \citenamefont {Santamarina},\ and\ \citenamefont {Mishra}}]{nauruzbayeva2020electrification}%
  \BibitemOpen
  \bibfield  {author} {\bibinfo {author} {\bibfnamefont {J.}~\bibnamefont {Nauruzbayeva}}, \bibinfo {author} {\bibfnamefont {Z.}~\bibnamefont {Sun}}, \bibinfo {author} {\bibfnamefont {A.}~\bibnamefont {Gallo~Jr}}, \bibinfo {author} {\bibfnamefont {M.}~\bibnamefont {Ibrahim}}, \bibinfo {author} {\bibfnamefont {J.~C.}\ \bibnamefont {Santamarina}}, \ and\ \bibinfo {author} {\bibfnamefont {H.}~\bibnamefont {Mishra}},\ }\bibfield  {title} {\enquote {\bibinfo {title} {Electrification at water--hydrophobe interfaces},}\ }\href@noop {} {\bibfield  {journal} {\bibinfo  {journal} {Nat. Commun.}\ }\textbf {\bibinfo {volume} {11}},\ \bibinfo {pages} {5285} (\bibinfo {year} {2020})}\BibitemShut {NoStop}%
\bibitem [{\citenamefont {Ratschow}\ \emph {et~al.}(2025)\citenamefont {Ratschow}, \citenamefont {Butt}, \citenamefont {Hardt},\ and\ \citenamefont {Weber}}]{ratschow2025liquid}%
  \BibitemOpen
  \bibfield  {author} {\bibinfo {author} {\bibfnamefont {A.~D.}\ \bibnamefont {Ratschow}}, \bibinfo {author} {\bibfnamefont {H.-J.}\ \bibnamefont {Butt}}, \bibinfo {author} {\bibfnamefont {S.}~\bibnamefont {Hardt}}, \ and\ \bibinfo {author} {\bibfnamefont {S.~A.~L.}\ \bibnamefont {Weber}},\ }\bibfield  {title} {\enquote {\bibinfo {title} {Liquid slide electrification: advances and open questions},}\ }\href@noop {} {\bibfield  {journal} {\bibinfo  {journal} {Soft Matter}\ }\textbf {\bibinfo {volume} {21}},\ \bibinfo {pages} {1251--1262} (\bibinfo {year} {2025})}\BibitemShut {NoStop}%
\bibitem [{\citenamefont {Lin}\ \emph {et~al.}(2020)\citenamefont {Lin}, \citenamefont {Xu}, \citenamefont {Wang},\ and\ \citenamefont {Wang}}]{lin2020quantifying}%
  \BibitemOpen
  \bibfield  {author} {\bibinfo {author} {\bibfnamefont {S.}~\bibnamefont {Lin}}, \bibinfo {author} {\bibfnamefont {L.}~\bibnamefont {Xu}}, \bibinfo {author} {\bibfnamefont {A.~C.}\ \bibnamefont {Wang}}, \ and\ \bibinfo {author} {\bibfnamefont {Z.~L.}\ \bibnamefont {Wang}},\ }\bibfield  {title} {\enquote {\bibinfo {title} {Quantifying electron-transfer in liquid-solid contact electrification and the formation of electric double-layer},}\ }\href@noop {} {\bibfield  {journal} {\bibinfo  {journal} {Nat. Commun.}\ }\textbf {\bibinfo {volume} {11}},\ \bibinfo {pages} {399} (\bibinfo {year} {2020})}\BibitemShut {NoStop}%
\bibitem [{\citenamefont {Li}\ \emph {et~al.}(2022)\citenamefont {Li}, \citenamefont {Bista}, \citenamefont {Stetten}, \citenamefont {Bonart}, \citenamefont {Sch{\"u}r}, \citenamefont {Hardt}, \citenamefont {Bodziony}, \citenamefont {Marschall}, \citenamefont {Saal}, \citenamefont {Deng}, \citenamefont {Berger}, \citenamefont {Weber},\ and\ \citenamefont {Butt}}]{li2022spontaneous}%
  \BibitemOpen
  \bibfield  {author} {\bibinfo {author} {\bibfnamefont {X.}~\bibnamefont {Li}}, \bibinfo {author} {\bibfnamefont {P.}~\bibnamefont {Bista}}, \bibinfo {author} {\bibfnamefont {A.~Z.}\ \bibnamefont {Stetten}}, \bibinfo {author} {\bibfnamefont {H.}~\bibnamefont {Bonart}}, \bibinfo {author} {\bibfnamefont {M.~T.}\ \bibnamefont {Sch{\"u}r}}, \bibinfo {author} {\bibfnamefont {S.}~\bibnamefont {Hardt}}, \bibinfo {author} {\bibfnamefont {F.}~\bibnamefont {Bodziony}}, \bibinfo {author} {\bibfnamefont {H.}~\bibnamefont {Marschall}}, \bibinfo {author} {\bibfnamefont {A.}~\bibnamefont {Saal}}, \bibinfo {author} {\bibfnamefont {X.}~\bibnamefont {Deng}}, \bibinfo {author} {\bibfnamefont {R.}~\bibnamefont {Berger}}, \bibinfo {author} {\bibfnamefont {S.~A.~L.}\ \bibnamefont {Weber}}, \ and\ \bibinfo {author} {\bibfnamefont {H-.J.}\ \bibnamefont {Butt}},\ }\bibfield  {title} {\enquote {\bibinfo {title} {Spontaneous charging affects the motion of sliding drops},}\ }\href@noop {} {\bibfield  {journal} {\bibinfo  {journal}
  {Nat. Phys.}\ }\textbf {\bibinfo {volume} {18}},\ \bibinfo {pages} {713--719} (\bibinfo {year} {2022})}\BibitemShut {NoStop}%
\bibitem [{\citenamefont {Singh}\ \emph {et~al.}(2025)\citenamefont {Singh}, \citenamefont {Ratschow}, \citenamefont {Aslam},\ and\ \citenamefont {Daniel}}]{singh2025bipolar}%
  \BibitemOpen
  \bibfield  {author} {\bibinfo {author} {\bibfnamefont {N.}~\bibnamefont {Singh}}, \bibinfo {author} {\bibfnamefont {A.~D.}\ \bibnamefont {Ratschow}}, \bibinfo {author} {\bibfnamefont {N.}~\bibnamefont {Aslam}}, \ and\ \bibinfo {author} {\bibfnamefont {D.}~\bibnamefont {Daniel}},\ }\bibfield  {title} {\enquote {\bibinfo {title} {Bipolar surface charging by evaporating water droplets},}\ }\href@noop {} {\bibfield  {journal} {\bibinfo  {journal} {arXiv preprint arXiv:2508.08884}\ } (\bibinfo {year} {2025})}\BibitemShut {NoStop}%
\bibitem [{\citenamefont {Wong}\ \emph {et~al.}(2011)\citenamefont {Wong}, \citenamefont {Kang}, \citenamefont {Tang}, \citenamefont {Smythe}, \citenamefont {Hatton}, \citenamefont {Grinthal},\ and\ \citenamefont {Aizenberg}}]{wong2011bioinspired}%
  \BibitemOpen
  \bibfield  {author} {\bibinfo {author} {\bibfnamefont {T.-S.}\ \bibnamefont {Wong}}, \bibinfo {author} {\bibfnamefont {S.~H.}\ \bibnamefont {Kang}}, \bibinfo {author} {\bibfnamefont {S.~K.~Y.}\ \bibnamefont {Tang}}, \bibinfo {author} {\bibfnamefont {E.~J.}\ \bibnamefont {Smythe}}, \bibinfo {author} {\bibfnamefont {B.~D.}\ \bibnamefont {Hatton}}, \bibinfo {author} {\bibfnamefont {A.}~\bibnamefont {Grinthal}}, \ and\ \bibinfo {author} {\bibfnamefont {J.}~\bibnamefont {Aizenberg}},\ }\bibfield  {title} {\enquote {\bibinfo {title} {Bioinspired self-repairing slippery surfaces with pressure-stable omniphobicity},}\ }\href@noop {} {\bibfield  {journal} {\bibinfo  {journal} {Nature}\ }\textbf {\bibinfo {volume} {477}},\ \bibinfo {pages} {443--447} (\bibinfo {year} {2011})}\BibitemShut {NoStop}%
\bibitem [{\citenamefont {Lafuma}\ and\ \citenamefont {Qu{\'e}r{\'e}}(2011)}]{lafuma2011slippery}%
  \BibitemOpen
  \bibfield  {author} {\bibinfo {author} {\bibfnamefont {A.}~\bibnamefont {Lafuma}}\ and\ \bibinfo {author} {\bibfnamefont {D.}~\bibnamefont {Qu{\'e}r{\'e}}},\ }\bibfield  {title} {\enquote {\bibinfo {title} {Slippery pre-suffused surfaces},}\ }\href@noop {} {\bibfield  {journal} {\bibinfo  {journal} {Europhys. Lett.}\ }\textbf {\bibinfo {volume} {96}},\ \bibinfo {pages} {56001} (\bibinfo {year} {2011})}\BibitemShut {NoStop}%
\bibitem [{\citenamefont {Lin}\ \emph {et~al.}(2025)\citenamefont {Lin}, \citenamefont {Wardani},\ and\ \citenamefont {Daniel}}]{Lin2025exploding}%
  \BibitemOpen
  \bibfield  {author} {\bibinfo {author} {\bibfnamefont {M.}~\bibnamefont {Lin}}, \bibinfo {author} {\bibfnamefont {F.}~\bibnamefont {Wardani}}, \ and\ \bibinfo {author} {\bibfnamefont {D.}~\bibnamefont {Daniel}},\ }\bibfield  {title} {\enquote {\bibinfo {title} {Exploding drops on lubricated surfaces},}\ }\href@noop {} {\bibfield  {journal} {\bibinfo  {journal} {Phys. Rev. Fluids}\ }\textbf {\bibinfo {volume} {``In press"}} (\bibinfo {year} {2025})}\BibitemShut {NoStop}%
\bibitem [{\citenamefont {Daniel}\ \emph {et~al.}(2017)\citenamefont {Daniel}, \citenamefont {Timonen}, \citenamefont {Li}, \citenamefont {Velling},\ and\ \citenamefont {Aizenberg}}]{daniel2017oleoplaning}%
  \BibitemOpen
  \bibfield  {author} {\bibinfo {author} {\bibfnamefont {D.}~\bibnamefont {Daniel}}, \bibinfo {author} {\bibfnamefont {J.~V.~I.}\ \bibnamefont {Timonen}}, \bibinfo {author} {\bibfnamefont {R.}~\bibnamefont {Li}}, \bibinfo {author} {\bibfnamefont {S.~J.}\ \bibnamefont {Velling}}, \ and\ \bibinfo {author} {\bibfnamefont {J.}~\bibnamefont {Aizenberg}},\ }\bibfield  {title} {\enquote {\bibinfo {title} {Oleoplaning droplets on lubricated surfaces},}\ }\href@noop {} {\bibfield  {journal} {\bibinfo  {journal} {Nat. Phys.}\ }\textbf {\bibinfo {volume} {13}},\ \bibinfo {pages} {1020--1025} (\bibinfo {year} {2017})}\BibitemShut {NoStop}%
\bibitem [{\citenamefont {Feng}\ \emph {et~al.}(2014)\citenamefont {Feng}, \citenamefont {Roch{\'e}}, \citenamefont {Vigolo}, \citenamefont {Arnaudov}, \citenamefont {Stoyanov}, \citenamefont {Gurkov}, \citenamefont {Tsutsumanova},\ and\ \citenamefont {Stone}}]{feng2014nanoemulsions}%
  \BibitemOpen
  \bibfield  {author} {\bibinfo {author} {\bibfnamefont {J.}~\bibnamefont {Feng}}, \bibinfo {author} {\bibfnamefont {M.}~\bibnamefont {Roch{\'e}}}, \bibinfo {author} {\bibfnamefont {D.}~\bibnamefont {Vigolo}}, \bibinfo {author} {\bibfnamefont {L.~N.}\ \bibnamefont {Arnaudov}}, \bibinfo {author} {\bibfnamefont {S.~D.}\ \bibnamefont {Stoyanov}}, \bibinfo {author} {\bibfnamefont {T.~D.}\ \bibnamefont {Gurkov}}, \bibinfo {author} {\bibfnamefont {G.~G.}\ \bibnamefont {Tsutsumanova}}, \ and\ \bibinfo {author} {\bibfnamefont {H.~A.}\ \bibnamefont {Stone}},\ }\bibfield  {title} {\enquote {\bibinfo {title} {Nanoemulsions obtained via bubble-bursting at a compound interface},}\ }\href@noop {} {\bibfield  {journal} {\bibinfo  {journal} {Nat. Phys.}\ }\textbf {\bibinfo {volume} {10}},\ \bibinfo {pages} {606--612} (\bibinfo {year} {2014})}\BibitemShut {NoStop}%
\bibitem [{\citenamefont {Keiser}\ \emph {et~al.}(2017)\citenamefont {Keiser}, \citenamefont {Bense}, \citenamefont {Colinet}, \citenamefont {Bico},\ and\ \citenamefont {Reyssat}}]{keiser2017marangoni}%
  \BibitemOpen
  \bibfield  {author} {\bibinfo {author} {\bibfnamefont {L.}~\bibnamefont {Keiser}}, \bibinfo {author} {\bibfnamefont {H.}~\bibnamefont {Bense}}, \bibinfo {author} {\bibfnamefont {P.}~\bibnamefont {Colinet}}, \bibinfo {author} {\bibfnamefont {J.}~\bibnamefont {Bico}}, \ and\ \bibinfo {author} {\bibfnamefont {E.}~\bibnamefont {Reyssat}},\ }\bibfield  {title} {\enquote {\bibinfo {title} {Marangoni bursting: evaporation-induced emulsification of binary mixtures on a liquid layer},}\ }\href@noop {} {\bibfield  {journal} {\bibinfo  {journal} {Phys. Rev. Lett.}\ }\textbf {\bibinfo {volume} {118}},\ \bibinfo {pages} {074504} (\bibinfo {year} {2017})}\BibitemShut {NoStop}%
\bibitem [{\citenamefont {Montanero}\ and\ \citenamefont {Gan{\'a}n-Calvo}(2020)}]{montanero2020dripping}%
  \BibitemOpen
  \bibfield  {author} {\bibinfo {author} {\bibfnamefont {J.~M.}\ \bibnamefont {Montanero}}\ and\ \bibinfo {author} {\bibfnamefont {A.~M.}\ \bibnamefont {Gan{\'a}n-Calvo}},\ }\bibfield  {title} {\enquote {\bibinfo {title} {Dripping, jetting and tip streaming},}\ }\href@noop {} {\bibfield  {journal} {\bibinfo  {journal} {Rep. Prog. Phys.}\ }\textbf {\bibinfo {volume} {83}},\ \bibinfo {pages} {097001} (\bibinfo {year} {2020})}\BibitemShut {NoStop}%
\bibitem [{\citenamefont {Wang}\ \emph {et~al.}(2020)\citenamefont {Wang}, \citenamefont {Zhang}, \citenamefont {Sun}, \citenamefont {Lin}, \citenamefont {Sun},\ and\ \citenamefont {Deng}}]{wang2020facile}%
  \BibitemOpen
  \bibfield  {author} {\bibinfo {author} {\bibfnamefont {Y.}~\bibnamefont {Wang}}, \bibinfo {author} {\bibfnamefont {W.}~\bibnamefont {Zhang}}, \bibinfo {author} {\bibfnamefont {Q.}~\bibnamefont {Sun}}, \bibinfo {author} {\bibfnamefont {S.}~\bibnamefont {Lin}}, \bibinfo {author} {\bibfnamefont {S.}~\bibnamefont {Sun}}, \ and\ \bibinfo {author} {\bibfnamefont {X.}~\bibnamefont {Deng}},\ }\bibfield  {title} {\enquote {\bibinfo {title} {Facile strategy to generate charged droplets with desired polarities},}\ }\href@noop {} {\bibfield  {journal} {\bibinfo  {journal} {ACS omega}\ }\textbf {\bibinfo {volume} {5}},\ \bibinfo {pages} {26908--26913} (\bibinfo {year} {2020})}\BibitemShut {NoStop}%
\bibitem [{\citenamefont {Yu}\ \emph {et~al.}(2025)\citenamefont {Yu}, \citenamefont {Ratschow}, \citenamefont {Tao}, \citenamefont {Li}, \citenamefont {Jin}, \citenamefont {Wang},\ and\ \citenamefont {Wang}}]{yu2025charged}%
  \BibitemOpen
  \bibfield  {author} {\bibinfo {author} {\bibfnamefont {F.}~\bibnamefont {Yu}}, \bibinfo {author} {\bibfnamefont {A.~D.}\ \bibnamefont {Ratschow}}, \bibinfo {author} {\bibfnamefont {R.}~\bibnamefont {Tao}}, \bibinfo {author} {\bibfnamefont {X.}~\bibnamefont {Li}}, \bibinfo {author} {\bibfnamefont {Y.}~\bibnamefont {Jin}}, \bibinfo {author} {\bibfnamefont {J.}~\bibnamefont {Wang}}, \ and\ \bibinfo {author} {\bibfnamefont {Z.}~\bibnamefont {Wang}},\ }\bibfield  {title} {\enquote {\bibinfo {title} {Why charged drops do not splash},}\ }\href@noop {} {\bibfield  {journal} {\bibinfo  {journal} {Phys. Rev. Lett.}\ }\textbf {\bibinfo {volume} {134}},\ \bibinfo {pages} {134001} (\bibinfo {year} {2025})}\BibitemShut {NoStop}%
\bibitem [{\citenamefont {Kreder}\ \emph {et~al.}(2018)\citenamefont {Kreder}, \citenamefont {Daniel}, \citenamefont {Tetreault}, \citenamefont {Cao}, \citenamefont {Lemaire}, \citenamefont {Timonen},\ and\ \citenamefont {Aizenberg}}]{kreder2018film}%
  \BibitemOpen
  \bibfield  {author} {\bibinfo {author} {\bibfnamefont {M.~J.}\ \bibnamefont {Kreder}}, \bibinfo {author} {\bibfnamefont {D.}~\bibnamefont {Daniel}}, \bibinfo {author} {\bibfnamefont {A.}~\bibnamefont {Tetreault}}, \bibinfo {author} {\bibfnamefont {Z.}~\bibnamefont {Cao}}, \bibinfo {author} {\bibfnamefont {B.}~\bibnamefont {Lemaire}}, \bibinfo {author} {\bibfnamefont {J.~V.~I.}\ \bibnamefont {Timonen}}, \ and\ \bibinfo {author} {\bibfnamefont {J.}~\bibnamefont {Aizenberg}},\ }\bibfield  {title} {\enquote {\bibinfo {title} {Film dynamics and lubricant depletion by droplets moving on lubricated surfaces},}\ }\href@noop {} {\bibfield  {journal} {\bibinfo  {journal} {Phys. Rev. X}\ }\textbf {\bibinfo {volume} {8}},\ \bibinfo {pages} {031053} (\bibinfo {year} {2018})}\BibitemShut {NoStop}%
\bibitem [{\citenamefont {Semprebon}\ \emph {et~al.}(2021)\citenamefont {Semprebon}, \citenamefont {Sadullah}, \citenamefont {McHale},\ and\ \citenamefont {Kusumaatmaja}}]{semprebon2021apparent}%
  \BibitemOpen
  \bibfield  {author} {\bibinfo {author} {\bibfnamefont {C.}~\bibnamefont {Semprebon}}, \bibinfo {author} {\bibfnamefont {M.~S.}\ \bibnamefont {Sadullah}}, \bibinfo {author} {\bibfnamefont {G.}~\bibnamefont {McHale}}, \ and\ \bibinfo {author} {\bibfnamefont {H.}~\bibnamefont {Kusumaatmaja}},\ }\bibfield  {title} {\enquote {\bibinfo {title} {Apparent contact angle of drops on liquid infused surfaces: geometric interpretation},}\ }\href@noop {} {\bibfield  {journal} {\bibinfo  {journal} {Soft Matter}\ }\textbf {\bibinfo {volume} {17}},\ \bibinfo {pages} {9553--9559} (\bibinfo {year} {2021})}\BibitemShut {NoStop}%
\bibitem [{\citenamefont {Dai}\ and\ \citenamefont {Vella}(2022)}]{dai_droplets_2022}%
  \BibitemOpen
  \bibfield  {author} {\bibinfo {author} {\bibfnamefont {Z.}~\bibnamefont {Dai}}\ and\ \bibinfo {author} {\bibfnamefont {D.}~\bibnamefont {Vella}},\ }\bibfield  {title} {\enquote {\bibinfo {title} {Droplets on lubricated surfaces: {The} slow dynamics of skirt formation},}\ }\href {\doibase 10.1103/PhysRevFluids.7.054003} {\bibfield  {journal} {\bibinfo  {journal} {Phys. Rev. Fluids}\ }\textbf {\bibinfo {volume} {7}},\ \bibinfo {pages} {054003} (\bibinfo {year} {2022})}\BibitemShut {NoStop}%
\bibitem [{\citenamefont {Guan}\ \emph {et~al.}(2015{\natexlab{b}})\citenamefont {Guan}, \citenamefont {Wells}, \citenamefont {Xu}, \citenamefont {McHale}, \citenamefont {Wood}, \citenamefont {Martin},\ and\ \citenamefont {Stuart-Cole}}]{guan2015evaporation}%
  \BibitemOpen
  \bibfield  {author} {\bibinfo {author} {\bibfnamefont {J.~H.}\ \bibnamefont {Guan}}, \bibinfo {author} {\bibfnamefont {G.~G.}\ \bibnamefont {Wells}}, \bibinfo {author} {\bibfnamefont {B.}~\bibnamefont {Xu}}, \bibinfo {author} {\bibfnamefont {G.}~\bibnamefont {McHale}}, \bibinfo {author} {\bibfnamefont {D.}~\bibnamefont {Wood}}, \bibinfo {author} {\bibfnamefont {J.}~\bibnamefont {Martin}}, \ and\ \bibinfo {author} {\bibfnamefont {S.}~\bibnamefont {Stuart-Cole}},\ }\bibfield  {title} {\enquote {\bibinfo {title} {{Evaporation of sessile droplets on slippery liquid-infused porous surfaces (SLIPS)}},}\ }\href@noop {} {\bibfield  {journal} {\bibinfo  {journal} {Langmuir}\ }\textbf {\bibinfo {volume} {31}},\ \bibinfo {pages} {11781--11789} (\bibinfo {year} {2015}{\natexlab{b}})}\BibitemShut {NoStop}%
\bibitem [{\citenamefont {Gelderblom}\ \emph {et~al.}(2022)\citenamefont {Gelderblom}, \citenamefont {Diddens},\ and\ \citenamefont {Marin}}]{gelderblom2022evaporation}%
  \BibitemOpen
  \bibfield  {author} {\bibinfo {author} {\bibfnamefont {H.}~\bibnamefont {Gelderblom}}, \bibinfo {author} {\bibfnamefont {C.}~\bibnamefont {Diddens}}, \ and\ \bibinfo {author} {\bibfnamefont {A.}~\bibnamefont {Marin}},\ }\bibfield  {title} {\enquote {\bibinfo {title} {Evaporation-driven liquid flow in sessile droplets},}\ }\href@noop {} {\bibfield  {journal} {\bibinfo  {journal} {Soft Matter}\ }\textbf {\bibinfo {volume} {18}},\ \bibinfo {pages} {8535--8553} (\bibinfo {year} {2022})}\BibitemShut {NoStop}%
\bibitem [{\citenamefont {Bian}\ \emph {et~al.}(2021)\citenamefont {Bian}, \citenamefont {Wang},\ and\ \citenamefont {McCarthy}}]{bian2021rediscovering}%
  \BibitemOpen
  \bibfield  {author} {\bibinfo {author} {\bibfnamefont {P.}~\bibnamefont {Bian}}, \bibinfo {author} {\bibfnamefont {Y.}~\bibnamefont {Wang}}, \ and\ \bibinfo {author} {\bibfnamefont {T.~J.}\ \bibnamefont {McCarthy}},\ }\bibfield  {title} {\enquote {\bibinfo {title} {Rediscovering silicones: the anomalous water permeability of “hydrophobic” pdms suggests nanostructure and applications in water purification and anti-icing},}\ }\href@noop {} {\bibfield  {journal} {\bibinfo  {journal} {Macromol. Rapid Commun.}\ }\textbf {\bibinfo {volume} {42}},\ \bibinfo {pages} {2000682} (\bibinfo {year} {2021})}\BibitemShut {NoStop}%
\bibitem [{\citenamefont {Giglio}\ \emph {et~al.}(2008)\citenamefont {Giglio}, \citenamefont {Gervais}, \citenamefont {Rangama}, \citenamefont {Manil}, \citenamefont {Huber}, \citenamefont {Duft}, \citenamefont {M{\"u}ller}, \citenamefont {Leisner},\ and\ \citenamefont {Guet}}]{giglio2008shape}%
  \BibitemOpen
  \bibfield  {author} {\bibinfo {author} {\bibfnamefont {E.}~\bibnamefont {Giglio}}, \bibinfo {author} {\bibfnamefont {B.}~\bibnamefont {Gervais}}, \bibinfo {author} {\bibfnamefont {J.}~\bibnamefont {Rangama}}, \bibinfo {author} {\bibfnamefont {B.}~\bibnamefont {Manil}}, \bibinfo {author} {\bibfnamefont {B.~A.}\ \bibnamefont {Huber}}, \bibinfo {author} {\bibfnamefont {D.}~\bibnamefont {Duft}}, \bibinfo {author} {\bibfnamefont {R.}~\bibnamefont {M{\"u}ller}}, \bibinfo {author} {\bibfnamefont {T.}~\bibnamefont {Leisner}}, \ and\ \bibinfo {author} {\bibfnamefont {C.}~\bibnamefont {Guet}},\ }\bibfield  {title} {\enquote {\bibinfo {title} {Shape deformations of surface-charged microdroplets},}\ }\href@noop {} {\bibfield  {journal} {\bibinfo  {journal} {Phys. Rev. E}\ }\textbf {\bibinfo {volume} {77}},\ \bibinfo {pages} {036319} (\bibinfo {year} {2008})}\BibitemShut {NoStop}%
\bibitem [{\citenamefont {Hunter}\ and\ \citenamefont {Ray}(2009)}]{hunter2009progeny}%
  \BibitemOpen
  \bibfield  {author} {\bibinfo {author} {\bibfnamefont {H.~C.}\ \bibnamefont {Hunter}}\ and\ \bibinfo {author} {\bibfnamefont {A.~K.}\ \bibnamefont {Ray}},\ }\bibfield  {title} {\enquote {\bibinfo {title} {On progeny droplets emitted during coulombic fission of charged microdrops},}\ }\href@noop {} {\bibfield  {journal} {\bibinfo  {journal} {Phys. Chem. Chem. Phys.}\ }\textbf {\bibinfo {volume} {11}},\ \bibinfo {pages} {6156--6165} (\bibinfo {year} {2009})}\BibitemShut {NoStop}%
\bibitem [{\citenamefont {Shumpert}(1972)}]{shumpert1972capacitance}%
  \BibitemOpen
  \bibfield  {author} {\bibinfo {author} {\bibfnamefont {T.~H.}\ \bibnamefont {Shumpert}},\ }\bibfield  {title} {\enquote {\bibinfo {title} {Capacitance calculations for satellites, part i. isolated capacitances of ellipsoidal shapes with comparisons to some other simple bodies},}\ }\href@noop {} {\bibfield  {journal} {\bibinfo  {journal} {Sensor and Simulation Notes}\ }\textbf {\bibinfo {volume} {157}},\ \bibinfo {pages} {1--30} (\bibinfo {year} {1972})}\BibitemShut {NoStop}%
\bibitem [{\citenamefont {Singh}\ \emph {et~al.}(2021{\natexlab{b}})\citenamefont {Singh}, \citenamefont {Gawande}, \citenamefont {Mayya},\ and\ \citenamefont {Thaokar}}]{singh2021subcritical}%
  \BibitemOpen
  \bibfield  {author} {\bibinfo {author} {\bibfnamefont {M.}~\bibnamefont {Singh}}, \bibinfo {author} {\bibfnamefont {N.}~\bibnamefont {Gawande}}, \bibinfo {author} {\bibfnamefont {Y.~S.}\ \bibnamefont {Mayya}}, \ and\ \bibinfo {author} {\bibfnamefont {R.}~\bibnamefont {Thaokar}},\ }\bibfield  {title} {\enquote {\bibinfo {title} {Subcritical asymmetric rayleigh breakup of a charged drop induced by finite amplitude perturbations in a quadrupole trap},}\ }\href@noop {} {\bibfield  {journal} {\bibinfo  {journal} {Phys. Rev. E}\ }\textbf {\bibinfo {volume} {103}},\ \bibinfo {pages} {053111} (\bibinfo {year} {2021}{\natexlab{b}})}\BibitemShut {NoStop}%
\bibitem [{\citenamefont {Lazo}\ and\ \citenamefont {Chen}(2024)}]{lazo2024self}%
  \BibitemOpen
  \bibfield  {author} {\bibinfo {author} {\bibfnamefont {J.~A.}\ \bibnamefont {Lazo}}\ and\ \bibinfo {author} {\bibfnamefont {R.-H.}\ \bibnamefont {Chen}},\ }\bibfield  {title} {\enquote {\bibinfo {title} {Self-similar behavior of successive coulombic fissions of evaporating charged water droplets},}\ }\href@noop {} {\bibfield  {journal} {\bibinfo  {journal} {Int. J. Heat Mass Transfer}\ }\textbf {\bibinfo {volume} {219}},\ \bibinfo {pages} {124879} (\bibinfo {year} {2024})}\BibitemShut {NoStop}%
\bibitem [{\citenamefont {Rayleigh}(1879)}]{rayleigh1879capillary}%
  \BibitemOpen
  \bibfield  {author} {\bibinfo {author} {\bibfnamefont {Lord}\ \bibnamefont {Rayleigh}},\ }\bibfield  {title} {\enquote {\bibinfo {title} {On the capillary phenomena of jets},}\ }\href@noop {} {\bibfield  {journal} {\bibinfo  {journal} {Proc. R. Soc. Lond.}\ ,\ \bibinfo {pages} {71--97}} (\bibinfo {year} {1879})}\BibitemShut {NoStop}%
\bibitem [{\citenamefont {Lamb}(1945)}]{lamb_hydrodynamics}%
  \BibitemOpen
  \bibfield  {author} {\bibinfo {author} {\bibfnamefont {H.}~\bibnamefont {Lamb}},\ }\href@noop {} {\emph {\bibinfo {title} {Hydrodynamics}}}\ (\bibinfo  {publisher} {Dover, New York},\ \bibinfo {year} {1945})\ \bibinfo {note} {6th ed., unabridged republication of the 1932 Cambridge edition}\BibitemShut {NoStop}%
\bibitem [{\citenamefont {Chang}\ \emph {et~al.}(2013)\citenamefont {Chang}, \citenamefont {Bostwick}, \citenamefont {Steen},\ and\ \citenamefont {Daniel}}]{chang2013substrate}%
  \BibitemOpen
  \bibfield  {author} {\bibinfo {author} {\bibfnamefont {C.-T.}\ \bibnamefont {Chang}}, \bibinfo {author} {\bibfnamefont {J.~B.}\ \bibnamefont {Bostwick}}, \bibinfo {author} {\bibfnamefont {P.~H.}\ \bibnamefont {Steen}}, \ and\ \bibinfo {author} {\bibfnamefont {S.}~\bibnamefont {Daniel}},\ }\bibfield  {title} {\enquote {\bibinfo {title} {Substrate constraint modifies the rayleigh spectrum of vibrating sessile drops},}\ }\href@noop {} {\bibfield  {journal} {\bibinfo  {journal} {Phys. Rev. E}\ }\textbf {\bibinfo {volume} {88}},\ \bibinfo {pages} {023015} (\bibinfo {year} {2013})}\BibitemShut {NoStop}%
\bibitem [{\citenamefont {Rubio}\ \emph {et~al.}(2023)\citenamefont {Rubio}, \citenamefont {Rodr{\'\i}guez-D{\'\i}az}, \citenamefont {L{\'o}pez-Herrera}, \citenamefont {Herrada}, \citenamefont {Ga{\~n}{\'a}n-Calvo},\ and\ \citenamefont {Montanero}}]{rubio2023role}%
  \BibitemOpen
  \bibfield  {author} {\bibinfo {author} {\bibfnamefont {M.}~\bibnamefont {Rubio}}, \bibinfo {author} {\bibfnamefont {P.}~\bibnamefont {Rodr{\'\i}guez-D{\'\i}az}}, \bibinfo {author} {\bibfnamefont {J.~M.}\ \bibnamefont {L{\'o}pez-Herrera}}, \bibinfo {author} {\bibfnamefont {M.~A.}\ \bibnamefont {Herrada}}, \bibinfo {author} {\bibfnamefont {A.~M.}\ \bibnamefont {Ga{\~n}{\'a}n-Calvo}}, \ and\ \bibinfo {author} {\bibfnamefont {J.~M.}\ \bibnamefont {Montanero}},\ }\bibfield  {title} {\enquote {\bibinfo {title} {The role of charge relaxation in electrified tip streaming},}\ }\href@noop {} {\bibfield  {journal} {\bibinfo  {journal} {Phys. Fluids}\ }\textbf {\bibinfo {volume} {35}} (\bibinfo {year} {2023})}\BibitemShut {NoStop}%
\bibitem [{\citenamefont {Collins}\ \emph {et~al.}(2013)\citenamefont {Collins}, \citenamefont {Sambath}, \citenamefont {Harris},\ and\ \citenamefont {Basaran}}]{collins2013universal}%
  \BibitemOpen
  \bibfield  {author} {\bibinfo {author} {\bibfnamefont {R.~T.}\ \bibnamefont {Collins}}, \bibinfo {author} {\bibfnamefont {K.}~\bibnamefont {Sambath}}, \bibinfo {author} {\bibfnamefont {M.~T.}\ \bibnamefont {Harris}}, \ and\ \bibinfo {author} {\bibfnamefont {O.~A.}\ \bibnamefont {Basaran}},\ }\bibfield  {title} {\enquote {\bibinfo {title} {Universal scaling laws for the disintegration of electrified drops},}\ }\href@noop {} {\bibfield  {journal} {\bibinfo  {journal} {Proc. Natl. Acad. Sci. U.S.A.}\ }\textbf {\bibinfo {volume} {110}},\ \bibinfo {pages} {4905--4910} (\bibinfo {year} {2013})}\BibitemShut {NoStop}%
\bibitem [{\citenamefont {Betel{\'u}}\ \emph {et~al.}(2006)\citenamefont {Betel{\'u}}, \citenamefont {Fontelos}, \citenamefont {Kindel{\'a}n},\ and\ \citenamefont {Vantzos}}]{betelu2006singularities}%
  \BibitemOpen
  \bibfield  {author} {\bibinfo {author} {\bibfnamefont {S.~I.}\ \bibnamefont {Betel{\'u}}}, \bibinfo {author} {\bibfnamefont {M.~A.}\ \bibnamefont {Fontelos}}, \bibinfo {author} {\bibfnamefont {U.}~\bibnamefont {Kindel{\'a}n}}, \ and\ \bibinfo {author} {\bibfnamefont {O.}~\bibnamefont {Vantzos}},\ }\bibfield  {title} {\enquote {\bibinfo {title} {Singularities on charged viscous droplets},}\ }\href@noop {} {\bibfield  {journal} {\bibinfo  {journal} {Phys. Fluids}\ }\textbf {\bibinfo {volume} {18}} (\bibinfo {year} {2006})}\BibitemShut {NoStop}%
\bibitem [{\citenamefont {Gan{\'a}n-Calvo}\ \emph {et~al.}(2016)\citenamefont {Gan{\'a}n-Calvo}, \citenamefont {L{\'o}pez-Herrera}, \citenamefont {Rebollo-Munoz},\ and\ \citenamefont {Montanero}}]{ganan2016onset}%
  \BibitemOpen
  \bibfield  {author} {\bibinfo {author} {\bibfnamefont {A.~M.}\ \bibnamefont {Gan{\'a}n-Calvo}}, \bibinfo {author} {\bibfnamefont {J.~M.}\ \bibnamefont {L{\'o}pez-Herrera}}, \bibinfo {author} {\bibfnamefont {N.}~\bibnamefont {Rebollo-Munoz}}, \ and\ \bibinfo {author} {\bibfnamefont {J.~M.}\ \bibnamefont {Montanero}},\ }\bibfield  {title} {\enquote {\bibinfo {title} {The onset of electrospray: the universal scaling laws of the first ejection},}\ }\href@noop {} {\bibfield  {journal} {\bibinfo  {journal} {Sci. Rep.}\ }\textbf {\bibinfo {volume} {6}},\ \bibinfo {pages} {32357} (\bibinfo {year} {2016})}\BibitemShut {NoStop}%
\bibitem [{\citenamefont {Stone}\ \emph {et~al.}(1999)\citenamefont {Stone}, \citenamefont {Lister},\ and\ \citenamefont {Brenner}}]{stone1999drops}%
  \BibitemOpen
  \bibfield  {author} {\bibinfo {author} {\bibfnamefont {H.~A.}\ \bibnamefont {Stone}}, \bibinfo {author} {\bibfnamefont {J.~R.}\ \bibnamefont {Lister}}, \ and\ \bibinfo {author} {\bibfnamefont {M.~P.}\ \bibnamefont {Brenner}},\ }\bibfield  {title} {\enquote {\bibinfo {title} {Drops with conical ends in electric and magnetic fields},}\ }\href@noop {} {\bibfield  {journal} {\bibinfo  {journal} {Proc. R. Soc. Lond. A}\ }\textbf {\bibinfo {volume} {455}},\ \bibinfo {pages} {329--347} (\bibinfo {year} {1999})}\BibitemShut {NoStop}%
\bibitem [{\citenamefont {Bradski}(2000)}]{opencv_library}%
  \BibitemOpen
  \bibfield  {author} {\bibinfo {author} {\bibfnamefont {G.}~\bibnamefont {Bradski}},\ }\bibfield  {title} {\enquote {\bibinfo {title} {{The OpenCV Library}},}\ }\href@noop {} {\bibfield  {journal} {\bibinfo  {journal} {Dr. Dobb's Journal of Software Tools}\ } (\bibinfo {year} {2000})}\BibitemShut {NoStop}%
\bibitem [{\citenamefont {Daniel}(2024)}]{dataset}%
  \BibitemOpen
  \bibfield  {author} {\bibinfo {author} {\bibfnamefont {D.}~\bibnamefont {Daniel}},\ }\href@noop {} {\enquote {\bibinfo {title} {{Replication dataset}},}\ }\bibinfo {howpublished} {\url{https://doi.org/10.7910/DVN/OFGOI3}} (\bibinfo {year} {2024})\BibitemShut {NoStop}%
\end{thebibliography}


\end{document}